\documentclass{article}

\usepackage[utf8]{inputenc}
\usepackage[margin=0.5in]{geometry}
\usepackage[titletoc,title]{appendix}
\usepackage{setspace}
\usepackage[dvipsnames]{xcolor}
\usepackage{hyperref}
\usepackage{color,soul}
\hypersetup{
  colorlinks,
  citecolor=blue,
  linkcolor=Black,
  urlcolor=blue}
\usepackage{indentfirst}
\usepackage{parskip}

\usepackage{amsmath,amsfonts,amssymb,mathtools}
\usepackage{authblk}
\usepackage{mathtools}

\DeclarePairedDelimiterX\braket[2]{\langle}{\rangle}{#1 \delimsize\vert #2}
\usepackage{graphicx,float}
\usepackage[font=small,labelfont=bf]{caption}
\usepackage{subcaption}
\usepackage[ruled,vlined]{algorithm2e}
\usepackage{algorithmic}

\newcommand{\np}[1]{n_\mathbf{p_#1}}
\newcommand{\NP}{n_\mathbf{p}}
\newcommand{\Vp}[3]{\left| V(\mathbf{p}_{#1}, \mathbf{p}_{#2}, \mathbf{p}_{#3})\right|^2}

\usepackage[sorting=none]{biblatex}
\bibliography{References}

\usepackage{enumitem}

\title{\textbf{Experimental Investigation of Tidally-Forced Internal Wave Turbulence at High Reynolds Number}}
\date{}

\author[1]{Zachary Taebel}
\author[2]{Alberto Scotti}
\author[3]{Pierre-Yves Passaggia}
\author[4]{Dylan Bruney}
\affil[1]{University of North Carolina, Chapel Hill, NC, USA}
\affil[2]{Arizona State University, Phoenix, AZ, USA}
\affil[3]{University of Orl\'{e}ans, Orl\'{e}ans, France}
\affil[4]{Wake Forest University, Winston-Salem, NC, USA}
\affil[ ]{ }
\affil[ ]{Email address for correspondence: taebel@live.unc.edu}

\begin{document}
\maketitle

\begin{abstract}
    \noindent Through basin-scale circulations, the ocean regulates global distributions of heat, nutrients, and greenhouse gases. To properly predict the future of the ocean under climate change, we need to develop a thorough understanding of the underlying mechanisms that drive global circulations. An estimated 2 TW of power is required to support interior mixing. Roughly half of this power is believed to come from tidal flow over topography, producing internal gravity waves (IGW's), which can radiate energy throughout the ocean interior. But it is difficult to track the subsequent journey from tidal injection to dissipation, as the energy cascade spans an enormous range of spatio-temporal scales and multiple different nonlinear transfer mechanisms. To investigate the full energy pathway from topographic forcing to irreversible mixing, we built a model ocean in a large-scale laboratory wavetank (9 m x 2.9 m x 0.75 m) allowing Reynolds numbers up to O(10$^5$). We replicate the tidal forcing by oscillating an idealized ocean ridge. We track energy transfer across the first cascade, driven by wave turbulence, using Background Oriented Schlieren (BOS) over the full tank. Through the BOS we observe the formation of various sets of subharmonics, driven by Triadic Resonant Instabilities (TRI). At later times, the subharmonics born from TRI engage in different interactions, which ultimately develop a continuum of waves at frequencies up to $N$. We validate the three-wave resonant conditions through a Fourier decomposition and confirm a backward cascade in frequency but a forward cascade in vertical wavenumber. Through our spatial analysis, we identify relevant three-wave interactions and show the significance of elastic scattering, a nonlocal interaction, in our fully evolved system. We note however that the majority of our triads are local, which have been historically overlooked. 
\end{abstract}

\section{Introduction}
The stratified waters of the ocean interior are an ideal environment for an internal wave field spanning a vast range of spatial and temporal scales. The smallest of these waves often become unstable and break, producing turbulence. These turbulent patches can do work against the local density gradients, and mix the water column. This wave-driven mixing can significantly govern local marine systems, such as replenishing nutrients that sustains life in the Mariana Trench \cite{van2020challenger} or driving overturns at the thermocline \cite{alford2000observations}. 

But its from a global standpoint that the true impact of wave-driven mixing becomes visible. The work of Munk \cite{munk1966abyssal} revealed that balancing deep-water formation with diapycnal upwelling in the global overturning circulation (GOC) requires diffusitives orders of magnitudes larger than simple molecular values. In short, the GOC can only survive with substantial internal mixing, which would require a 2 terawatt (TW) power source \cite{wunsch2004vertical}.  

Satellite altimetry \cite{egbert2000significant} predict roughly half of this required power for mixing ($\approx$ 1 TW) comes from barotropic tides that dissipate in the abyssal ocean \cite{garrett2003internal,garrett2007internal}. Observations and slab models show that roughly the other half can be attributed to oscillations of the mixed layer driven by wind flow \cite{whalen2020internal}. This is where internal waves enter the picture. The energy supplied by tidal flow and wind-driven oscillations of the mixed layer is transferred into internal waves (internal waves produced by tidal flow are termed internal tides), which radiate away from their generation site and carry energy into the ocean interior \cite{mackinnon2013diapycnal}. Wave breaking then uses the supplied energy to produce the required mixing. 

But this simple view overlooks the complexity of the mechanisms by which internal waves can move energy from large injection scales down to small dissipative scales \cite{staquet2002internal}. Tidal forcing over ocean ridges can excite waves at a multitude of spatial scales, but only the highest modes at vertical wavelengths of $O$(10m) or less are likely to break via shear instability \cite{Laurent2002}. An estimated 70\% of the tidal energy converted to internal tides is radiated into the far field as low-mode waves \cite{StLaurent2002Tidal}, at length scales still well removed from overturning and mixing scales \cite{Laurent2002, FoxKemper2019}. 

Interactions with topography dissipate some of the low-mode internal tides \cite{Laurent2002}, but 65-80\% of the energy dissipation of low-modes is attributed to wave-wave interactions \cite{whalen2020internal, deLavergne2019}. Assuming a slow nonlinear timescale, these interactions arise from plugging the linear wave solutions into the fully nonlinear equations of motion. The resulting terms have phases at the sums and differences of the original waves, and provided they meet a series of resonance conditions, they represent new linear waves produced via the interactions \cite{tanaka2022physics}. Through these interactions, energy is moved to new spatial and temporal scales.  The statistical description of these interactions in a system of many random linear waves is termed wave turbulence (WT). 

An objective of WT is to derive a single equation for the evolution of the wave action spectrum $n_\mathbf{p}$, termed the wave kinetic equation (WKE). This approach dates back to Hasselmann \cite{hasselmann1962non} for surface waves, but has since been applied to a variety of nonlinear systems \cite{nazarenko2011wave}. For internal waves, the derivation generally starts from isopyncnal coordinates and a canonical Hamiltonian framework \cite{lvov2005scale}, and requires assumptions of weak nonlinearity and Gaussian statistics in wavenumber space \cite{lvov2012resonant}. Note that there are other routes to the WKE involving Eulerian \cite{lvov2001hamiltonian,lvov2004hamiltonian} and Lagrangian coordinates \cite{olbers1974energy,olbers1976nonlinear}, as well as non-Hamiltonian formulations \cite{caillol2000kinetic}. Fortunately, it has been shown by Lvov \textit{et al.} \cite{lvov2012resonant} that in the absence of background rotation, all derivations lead to kinetic equations which are dynamically equivalent on the resonant manifold. We thus present the standard form of the WKE for internal waves, written as \cite{zakharov2012kolmogorov}\begin{equation}
\begin{split}
   \frac{\partial \NP}{\partial t} &= \iint 4\pi\Vp{}{1}{2}\left(\np{1}\np{2}-\NP\np{1}-\NP\np{2}\right)\delta(\mathbf{p}-\mathbf{p}_1-\mathbf{p}_2)\delta(\omega_\mathbf{p}-\omega_{\mathbf{p}_1}-\omega_{\mathbf{p}_2})\mathrm{d}\mathbf{p}_1\mathrm{d}\mathbf{p}_2\\
   &- \iint 8\pi\Vp{1}{}{2}\left(\NP\np{2}-\NP\np{1}-\np{1}\np{2}\right)\delta(\mathbf{p}_1-\mathbf{p}-\mathbf{p}_2)\delta(\omega_{\mathbf{p}_1}-\omega_{\mathbf{p}}-\omega_{\mathbf{p}_2})\mathrm{d}\mathbf{p}_1\mathrm{d}\mathbf{p}_2
   \end{split} 
\end{equation}
where the term $V(\mathbf{p},\mathbf{p}_1,\mathbf{p}_2)$ is an interaction coefficient, defined in \cite{lvov2001hamiltonian}, and $\mathbf{p} = (k_x,k_y,m) = (\mathbf{k},m)$ is the three-dimensional wavevector.  The entire right-hand side, termed the collision integral, represents how the action density evolves through every possible triad of waves in resonance with waves at wavevector $\mathbf{p}$. These resonances are imposed through the delta functions, which mandate that triads must simultaneously obey the following two relationships \cite{pan2020numerical}: 
\begin{align}
    \mathbf{p} &= \mathbf{p}_1 \pm \mathbf{p}_2 \\ 
    \omega_\mathbf{p} &= \omega_{\mathbf{p}_1} \pm \omega_{\mathbf{p}_2}
\end{align}
That these conditions must be simultaneously satisfied inherently requires all modes to obey the dispersion relation
\begin{equation}
    \omega^2_\mathbf{p} = \frac{N^2k^2+f^2m^2}{m^2 + k^2},
\end{equation}
meaning they are all linear waves \cite{taebel2022investigation}. The waves that satisfy these conditions define the aforementioned resonant manifold of the WKE \cite{nazarenko2011wave}. In (4), $N$ is the Brunt-V$\ddot{\mathrm{a}}$is$\ddot{\mathrm{a}}$l$\ddot{\mathrm{a}}$ (BV) frequency, $f$ is the Coriolis frequency, and $k = |\mathbf{k}|$.   

\begin{figure}[h]
    \centering
    \includegraphics[width=0.6\textwidth]{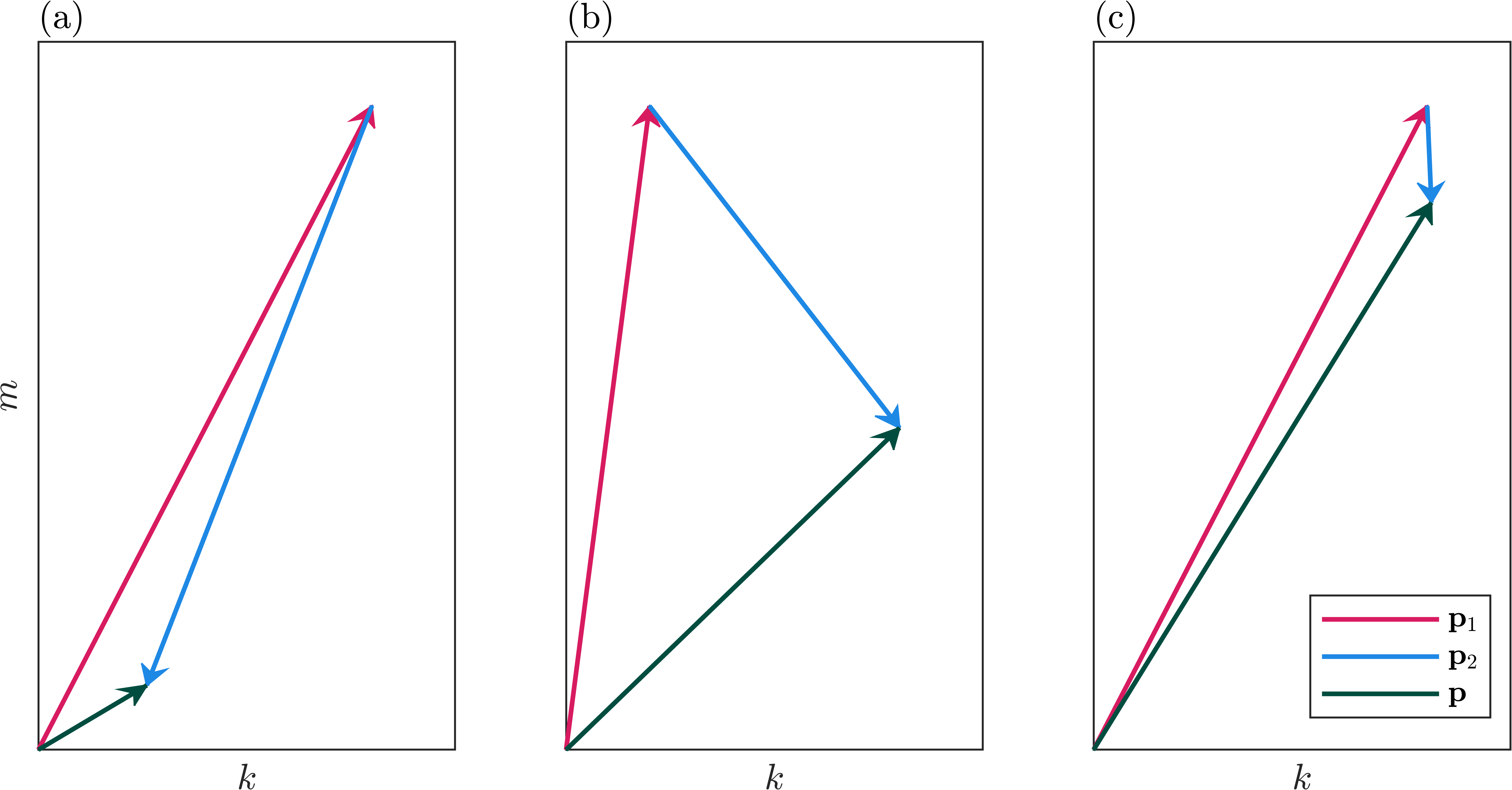}
    \caption{Example triads of wavevectors $(\mathbf{p}.\mathbf{p}_1,\mathbf{p}_2)$ satisfying both (2) and (3), and having the nonlocal character of (a) PSI, (b) ES, (c) ID. $m$ and $k$ are the vertical and horizontal wavenumber, respectively.}
    \label{fig:nonlocaltriads}
\end{figure}

The motivation for applying WT and the WKE to internal waves largely stems from the semi-empirical Garret-Munk (GM) spectrum \cite{garrett1972space,garrett1975space}. This spectrum was developed assuming linear internal waves, and fit to observational data. At high frequencies and high wavenumbers, the GM spectrum asymptotically behaves as a power law $\omega^{-2}m^{-2}$. Though deviations exist near boundaries and the equator \cite{polzin2011toward}, the spectrum has shown remarkable consistency with observations \cite{cushman2011introduction} and motivated the notion of a ``universal" background internal wave spectrum \cite{wunsch2018100,lvov2001hamiltonian, briscoe1975internal}. But the utility of GM in describing what is happening belies its inability to describe why it is happening. Being empirical, it doesn't reveal the underlying physics that produce the observed spectra \cite{allen1989statistical}, and this is where WT enters the picture. Since GM is based on linear internal waves, it stands to reason that the spectral development is at least partially driven by wave resonances \cite{lvov2012resonant}. Additionally, WT predicts a forward cascade of energy \cite{nazarenko2011wave} with corresponding power law spectral solutions to the WKE in the limit of large $\omega$ and large $m$ \cite{polzin2011toward}. Unsurprisingly, many authors have attempted to use WT and the WKE to explain oceanic spectra, as well as downscale energy transfers from large injection scales. Perhaps the most impactful of these studies was that of McComas and Bretherton (MB) \cite{mccomas1977resonant}. They suggested that three special groups of interactions on the resonant manifold dominate the collision integral. These groups of interactions are termed parametric subharmonic instability (PSI), elastic scattering (ES), and induced diffusion (ID). These interactions are all nonlocal (in frequency and/or wavenumber space), and allow considerable simplifications of the WKE \cite{mccomas1981time}.  A diagram of the structure of the nonlocal triads is shown in Fig. \ref{fig:nonlocaltriads}, taken from \cite{wu2023energy}. But despite the promise that the simplified MB model presents, the significance of these nonlocal interactions has recently been called into question \cite{dematteis2023structure, wu2023energy}. 

Of the three nonlocal interactions, PSI has by far the most observational \cite{xie2011observations, xie2013observations, alford2008observations, liu2020disintegration}, numerical \cite{mackinnon2005subtropical, wang2021resonant, richet2018internal, gayen2013degradation}, and experimental \cite{clark2010generation, joubaud2012experimental, bourget2013experimental, dossmann2017mixing} evidence. This is unsurprising since PSI arises from an inherent instability of monochromatic internal waves and exists outside the purview of WT \cite{sarkar2017topographic}. The work by Bourget \textit{et al.} \cite{bourget2014finite} demonstrated that the instability can also occur for internal wave beams, which are monochromatic in frequency but not in wavenumber \cite{sutherland2013wave, sutherland2010internal}. Consequently, PSI of the internal tidal beams has received considerable focus in ocean simulations for it's ability to induce local mixing through elevated shear and downscale energy transfers \cite{wang2021resonant, mackinnon2005subtropical}. This process is nonresonant poleward of the critical latitude, where the local Coriolis frequency equals half of the tidal frequency. Within some of these works, the subharmonics formed are not at or near half of the driving frequency, whereas PSI in an ideal fluid leads to daughter waves at half of the driving frequency. This is a consequence of viscosity, which modulates the growth rate of the subharmonics such that the peak growth rate is not at half of the original frequency \cite{bourget2013experimental}. As a result, it is common to use triadic resonant instability (TRI) to refer to the broader class of three-wave instabilities which are reminiscent of PSI, but incorporate viscous effects. TRI is not included in the MB model, and it perhaps cannot be, since the theory of WT typically assumes the dissipation timescale to sufficiently larger than the nonlinear timescale that viscous effects are neglected \cite{Campagne2018}. In some oceanic situation, the frequencies of TRI may also be influenced by internal wave breaking, which tends to introduce near-inertial waves \cite{nikurashin2011mechanism}.

An important consequence of PSI/TRI is the inevitable introduction of new waves within the system. This increases the number of possible triads, particularly since the daughter waves can become parent waves to their own instability. The development of many new waves and new triads leads the system towards a more fully WT state \cite{korobov2008interharmonics}. This phenomenon is well observed in prior experimental studies on internal WT. In particular, the work of Davis \textit{et al.} \cite{davis2020succession} using a wave attractor demonstrates how successive TRI interactions result in a roughly continuous spectrum in frequency and a power law in wavenumber. This work is consistent with non-attractor studies in the Coriolis facility by Rodda \textit{et al.} \cite{rodda2022experimental} and Savaro \textit{et al.} \cite{savaro2020generation}. In these experiments the authors found the early stages of energy transfer to be dominated by PSI/TRI, but once enough daughter waves had formed they identified triads between different sets of daughter waves. This caused the system to tend to a smoother spectrum. Further work by Lanchon \textit{et al.} \cite{lanchon2023internal} showed that an ever richer and more continuous regime of WT could be attained by altering the setup to avoid formation of modal waves, those being the fundamental eigenmodes of the domain. 

In contrast to the attention on PSI/TRI, the role of ID and ES in internal WT remains largely unexplored outside of more theoretical studies. The significance of these interactions has some corroboration in numerical simulations \cite{pan2020numerical}, but no evidence experimentally. This is an extremely salient question given the recent work of Dematteis \textit{et al.} \cite{dematteis2022origins, dematteis2023structure} and Wu and Pan \cite{wu2023energy}. In these works, the authors challenge the idea set forth by McComas and Bretherton that the WKE is dominated by the three special classes of scale-separated interactions. A huge conclusion that all three aforementioned studies agree on is that local interactions, neglected in the MB model, are just as important as the nonlocal interactions. But there is disagreement between Dematteis \textit{et al.} \cite{dematteis2022origins, dematteis2023structure} and Wu and Pan \cite{wu2023energy} regarding which of the nonlocal interactions are most significant. Dematteis \textit{et al.} \cite{dematteis2022origins, dematteis2023structure} found the collision integral to be dominated by ID and local interactions for the true power law solution to the WKE, which may be seen as the expected behavior of internal WT. Wu and Pan \cite{wu2023energy} instead considered the GM spectrum and similarly identified the dominance of local interactions, but they identified ES and PSI as the dominant nonlocal interactions. ID was not only subdominant but as illustrated in Fig. 1 of \cite{wu2023energy}, had a variable preferential direction of action transfer. These studies have made it evident that there is an urgency to validate whether the principals of WT can be applied to a real world system if we are to recognize it within the larger model of the oceanic energy cascade. 

With all this in mind we identify two major questions we seek to answer in our study. The first is whether we can corroborate WT as the first stage of energy transfer from large injection scales down to smaller overturning scales through a laboratory experiment. The work of Kunze \cite{kunze2019unified} showed that it would be a mistake to label WT as the sole mechanism of energy transfer to mixing scales, and thus we seek only a sufficient bridge to kick-start further nonlinear dynamics. This has been studied previously by Brouzet \textit{et al.} \cite{brouzet2016energy} for internal wave attractors and by Rodda \textit{et al.} \cite{rodda2023internal} for waves in the Coriolis facility. Both studies identified regions of elevated stratified turbulence, overturning, and mixing driven at large scales by WT. The problem with these studies however is that from an oceanographic perspective, the forcing conditions are unrealistic. Internal wave attractors are an artifact of specific geometries and have not been observed in nature, even in a double ridge system \cite{manders2004observations}. Additionally, the linear focusing of energy in attractors leads to concentrated elevated dissipation along the beam rather than in the fluid background \cite{brouzet2016internal}, which is unlikely to be responsible for global ocean interior mixing. The work at the Coriolis facility in contrast uses an inherently modal wave forcing, rather than a wave beam formed by tidal flow over topography. An important point drawn by the prior studies \cite{rodda2022experimental,rodda2023internal,savaro2020generation} is the sensitivity of the resulting WT to the forcing parameters. Thus one can not immediately assume the results hold true under a different forcing mechanism. But the need for a tidal forcing is also relevant given the work of Chen \textit{et al.} \cite{chen2019can}, who presented numerical results suggesting that through only a tidal forcing one can trigger a GM spectral cascade.

The second question is what types of triad are most notably involved in the spectral development and energy transfer in our system. This questions stands to partially address the current disagreements regarding the importance of nonlocal interactions, as well as the role of the historically ignored local triads. From prior work we expect PSI/TRI to be important in early development of a WT system, but it is unclear what role other triads will have in developing a more fully turbulent system. It is important to note that in an experimental approach to internal waves, it is difficult to have a large separation between the forcing frequency $\omega_0$ and $N$ while still having a sufficiently large Reynolds number for nonlinear effects \cite{rodda2022experimental}. Thus our investigation does not seek to address the formation of a high frequency power law such as GM, but rather considers the behaviors predominantly below $\omega_0$. In the ocean the subharmonic behavior is often exclusively PSI/TRI, resulting in isolated peaks. However locations closer to the equator which have larger separation between $f$ and the tidal frequency are reminiscent of the regimes explored in past laboratory experiments, and indeed some spectral results show energy content distributed across many more frequencies than just PSI peaks, such as Fig. 9(a) from Liao \textit{et al.} \cite{liao2012current}.

We therefore design an experimental study on internal wave turbulence forced via realistic tidal flow. We perform our experiment in a large wavetank to achieve a sufficiently large Reynolds number, and observe a turbulent cascade starting from just a monochromatic tidal forcing. The present work uses a deeper large-scale facility than past experiments, leaving room for a wider range of temporal and spatial scales. We find low frequencies to be an important factor in driving WT. At early times these low frequencies are formed through multiple stages of PSI/TRI, which sets the stage for a greater panoply of interactions to occur. At later times, we observe a continuous spectrum in frequency, with large energy content at extremely low modes. Through higher order spectral methods, we show that our spectrum is composed of linear internal waves, and confirm a downscale energy transfer driven by WT. Importantly, we also observe significant effects of elastic scattering and local interactions in our system, providing the first experimental evidence to corroborate the work of Dematteis \textit{et al.} \cite{dematteis2022origins, dematteis2023structure} and Wu and Pan \cite{wu2023energy}. Our work provides strong support for WT as a first stage of downscale energy transfer from tidal injection scales.    

\section{Methods}

\begin{figure}[h]
    \centering
    \includegraphics[width=0.5\textwidth]{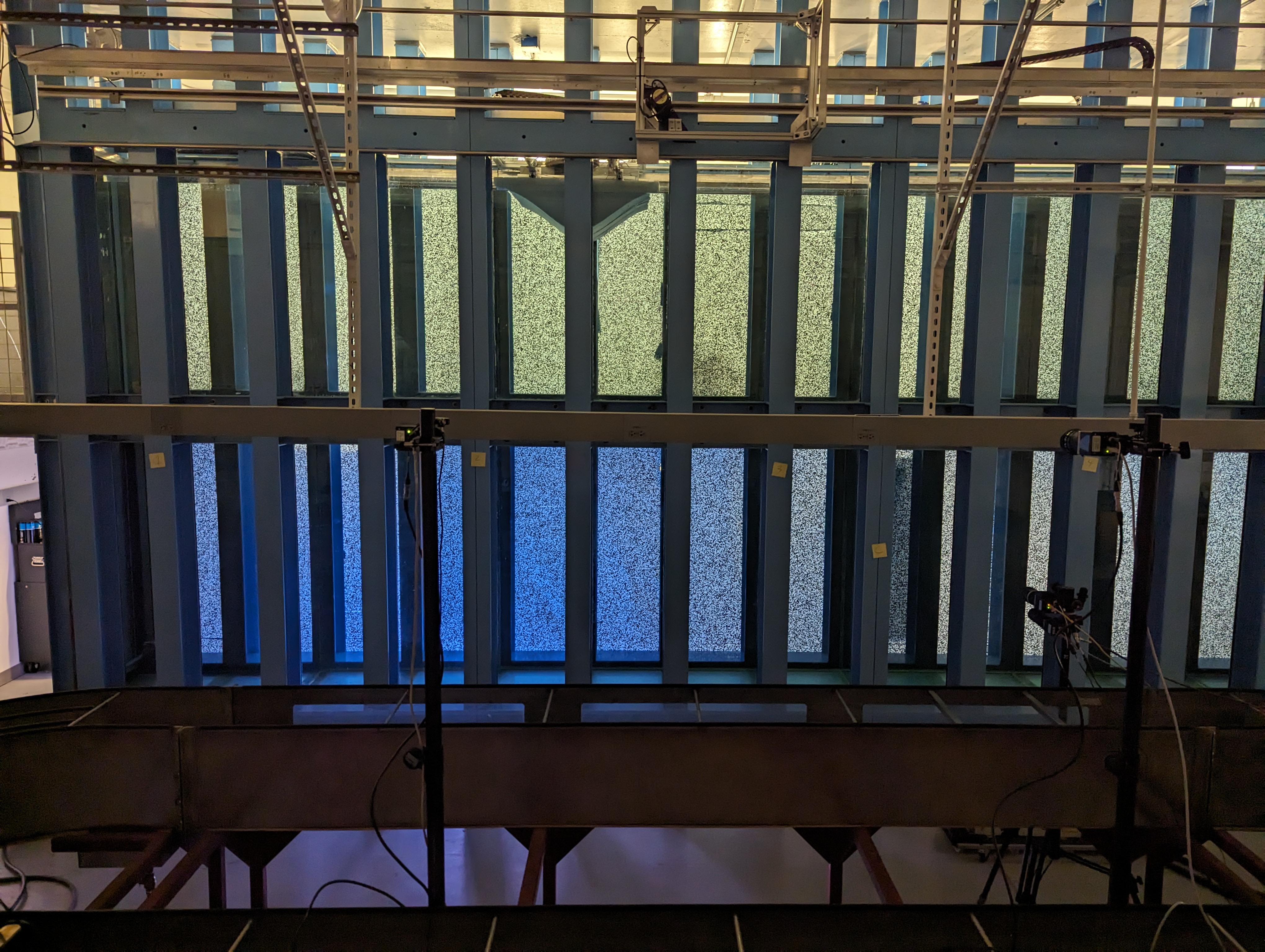}
    \caption{Photo of experimental setup showing random dot pattern paper on the far side of the tank, illuminated by lights.}
    \label{fig:labphoto}
\end{figure}

We performed experiments at the Joint Applied Mathematics and Marine Sciences Fluids Lab at UNC Chapel Hill. The facility features a large rectangular wavetank 36 meters in length, composed of three sections which can be isolated via metal gates. We worked in the deep section of the tank, which is 9 m long, 3 m tall, and 0.75 m wide. A photo of the deep section is shown in Fig. \ref{fig:labphoto}. The walls of the tank consist of regularly spaced glass windows in a grid pattern, supported by a metal frame. Each window is 0.42 m wide by 1.47 m tall, although the leftmost and rightmost windows are slightly narrower.

\begin{figure}[h]
    \centering
    \includegraphics[width=0.5\textwidth]{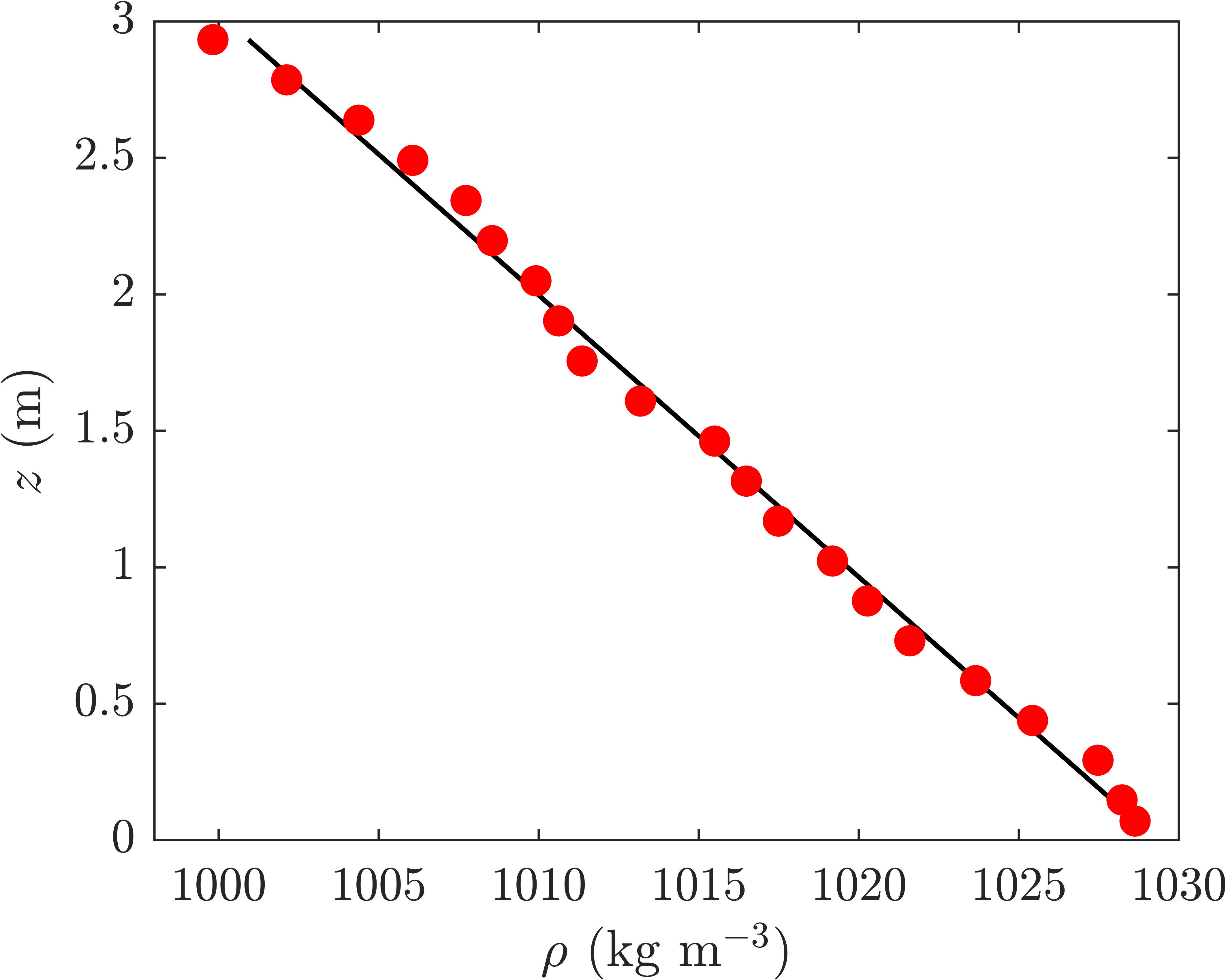}
    \caption{Vertical profile of fluid density $\rho$ vs height ($z$) prior to the experiment. The density was measured using an Orion Star conductivity probe, and converted to density via calibration. The black line denotes the linear fit for $\rho(z)$, with corresponding BV frequency $N = 0.309$ rad s$^{-1}$.}
    \label{fig:stratification}
\end{figure}

We stratified the tank with the two bucket method \cite{oster1963density} using a saltwater facility, and filled the tank to a depth of 2.93 m. Our stratification is approximately linear, in that the density does not vary continuously in the vertical, but rather is composed of many thin homogeneous layers. These layers have vertical thickness $O$(cm), which is an order of magnitude smaller than the vertical scales of the forcing and the waves. Additionally, the density values of each layer are chosen such that our discretized values of $\rho$ are fit by a linear function in $z$. This is corroborated by our measured density profile before the experiment, shown in Fig. \ref{fig:stratification}. We obtained these measurements using an Orion Star A215 conductivity meter, connected to a moveable cart that slid along a vertical rail near the right end of the tank before the experiment. We calibrated the probe using varying salinity water, and measured density at 21 depths which were close to uniformly spaced. Based on our linear fit, we assign a BV frequency $N = 0.309$ rad s$^{-1}$.

\begin{figure}[h]
    \centering
    \includegraphics[width=0.64\textwidth]{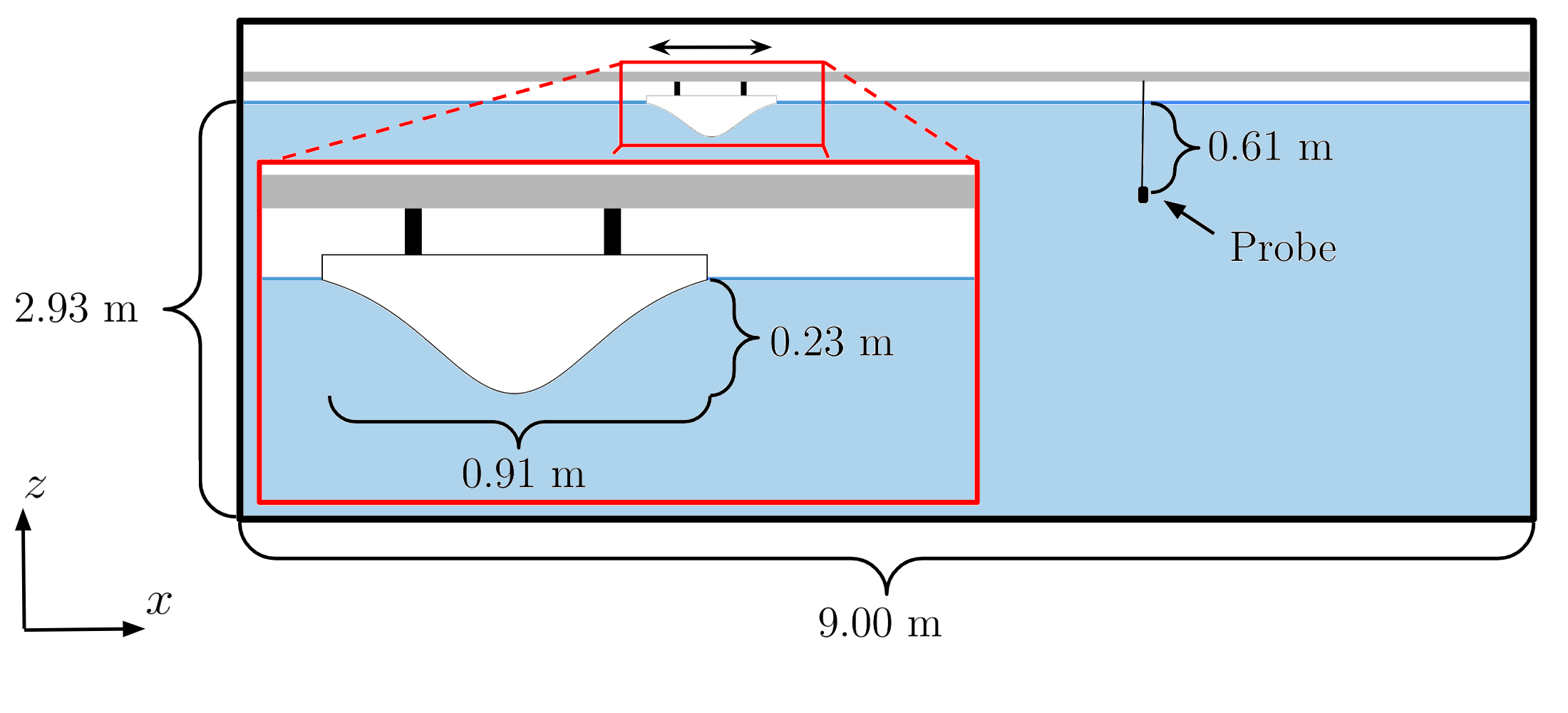}
    \includegraphics[width=0.35\textwidth]{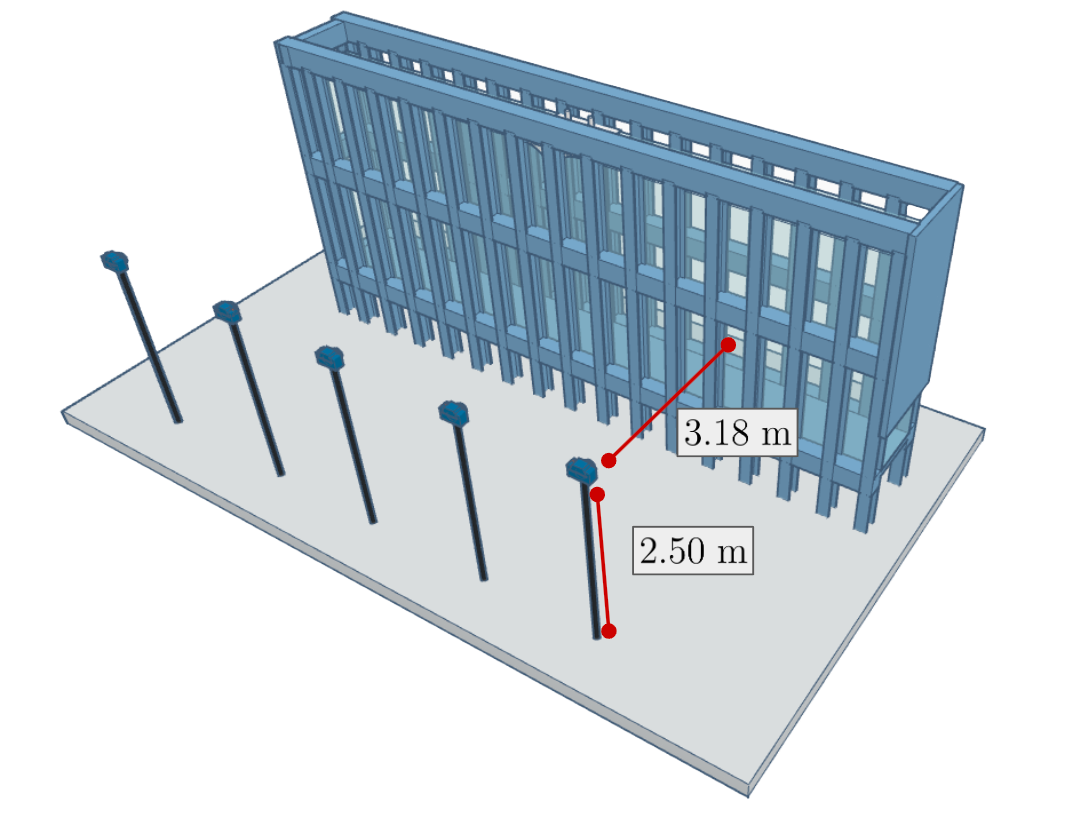}
    \caption{(Left) Schematic of experimental setup within the tank. Topography is mounted upside-down and oscillated about the mean position shown above. The position of the PME temperature-conductivity probe is shown. Note that the metal frame of the tank is excluded from the schematic for simplicity. (Right) 3D model of the experimental setup showing camera arrangement for BOS.}
    \label{fig:tank2d}
\end{figure}

Within the tank, we forced the system using an idealized model ocean ridge. Our topography was a Styrofoam block cut in the shape of a hyperbolic secant squared, defined as 
\begin{equation*}
    h(x) = h_0\,\mathrm{sech}^2(x/l_0) - c_0,
\end{equation*}
with maximum height $h_0 = 0.27$ m, length parameter $l_0$ = 0.27 m, and vertical shift $c_0$ = 0.04 m. The secant profile sat on top of a 0.09 m thick layer of uncut Styrofoam to serve as a base. We submerged the ridge upside-down at a depth of 0.27 m, such that the full topographic length $L = 1.143$ m. Within a Boussinesq approximation, an upside-down oscillating ridge is dynamically equivalent to a right-side-up stationary ridge with oscillating fluid via a change of reference frame \cite{spiegel1960boussinesq}. This setup is commonly used in experiments on internal tides for convenience \cite{echeverri2009low, lee2019turning, lee2020evanescent, aguilar2006internal, aguilar2006laboratory}. We suspended the topography from a movable cart which slides along horizontal rails at the top of the tank. We then provide an oscillatory forcing using a stepper motor attached to a linear stage sitting on the top of the tank (see for a detailed description of the facility \cite{camassa2018experimental}) by prescribing a discrete set of positions with respect to time, taken from a cosine function $x(t) = A\cos(\omega_0t)$. Oscillations have an amplitude $A = 0.41$ m and a frequency $\omega_0 = 0.26$ rad s$^{-1}$. To compensate for the upwards buoyant force the water applies to the topography once submerged, we slowly added 90 kg of weight to the top of the cart during the stratification process. The width of the topography was 0.74 m, which is slightly less than the width of the tank. Therefore the system can be considered 2D at large scales. A schematic of our forcing setup is shown in Fig. \ref{fig:tank2d}. Note that the mean position of our topography is offset from the center of the tank to avoid the formation of an attractor. 

We captured these large-scale wave dynamics performing background-oriented schlieren (BOS), also known as synthetic schlieren. Five GigE Bobcat B3320 Imperx 8Mpx CCD cameras, each individually filmed a $2\times3$ grid of six windows at a sampling rate of $1$ Hz. Simultaneous synchronization between all cameras was ensured using a Stanford Research DG535 pulse generator. The camera arrangement is shown in Fig. \ref{fig:tank2d}(b). 
%\R{PIERRE I THINK I'LL NEED YOU TO FILL IN THE REST OF THIS PART OF THE METHODS}.
Density gradients were then directly obtained from the difference between images acquired during the experiment and an image taken at rest, before the start of the experiment and a couple hours after the stratification was poured. Data processing followed using the cross-correlation method described in \cite{passaggia2020estimating}. In this particular type of BOS setup, the major issue is to ensure that pixel displacement is within the range acceptable for particle image velocimetry (i.e. 5 to 10 pixels). All cameras being identical and at a similar distance from the tank, the minimum displacement is set by the mean (undisturbed) vertical density gradient, measured between a snapshot taken with the tank filled with fresh water and the undisturbed stratification. The resulting mean vertical displacement was $1.2$ px while displacement during the experiment ranged between $3$ to $9$ px. Calibration was performed using the data obtained from the salinity probe and the vertical gradient obtained from the BOS. The metal frame of the tank hides enough of the field of view to make it difficult to solve for the density (and thus, buoyancy) field, and so we work with the gradient of buoyancy. 

% \begin{figure}[h]
%     \centering
%     \includegraphics[width=0.5\textwidth]{TankV@NoBackground.png}
%     \caption{3D model of experimental setup showing camera arrangement for BOS}
%     \label{fig:tank3d}
% \end{figure}

To supplement the global picture from the BOS with local microstructure measurements, we took continuous recordings from a PME temperature-conductivity probe at a sampling rate of 10 Hz. We submerged the probe 61 cm below the surface, at a location where the wave at frequency $\omega_0$ will propagate after reflection at the bottom of the tank. We calibrated the probe using four samples of water at fixed temperature and varying salinity.  

We define the relevant dimensionless numbers for our system, based on our choice of experimental parameters. The excursion number $Ex = 2A/L = 0.9$ and frequency ratio $\omega_0/N = 0.84$ position us in a nonlinear and nonhydrostatic regime, where secondary effects at the site of generation such as lee waves and overturns are likely to occur. However, this choice of parameters sets our Reynolds number $Re$ at $O(10^5)$. From prior experiments at smaller scales and numerical simulations, we predict that this value of $Re$ should be sufficient to enable wave-driven nonlinear dynamics. Nevertheless, we compensated for an overly nonlinear forcing by using a subcritical topography, with a ratio of 0.66 between the maximum topographic slope and the phase lines of the waves. 

The experimental duration was 131 minutes of active forcing, with an additional ten minutes of measurement after the tidal oscillation ends to record the decay of the wave field. We also recorded the stratification after the experiment to determine irreversible mixing caused by the waves. 

\section{Results}

\begin{figure}[h!]
    \centering
    \includegraphics[width=0.7\textwidth]{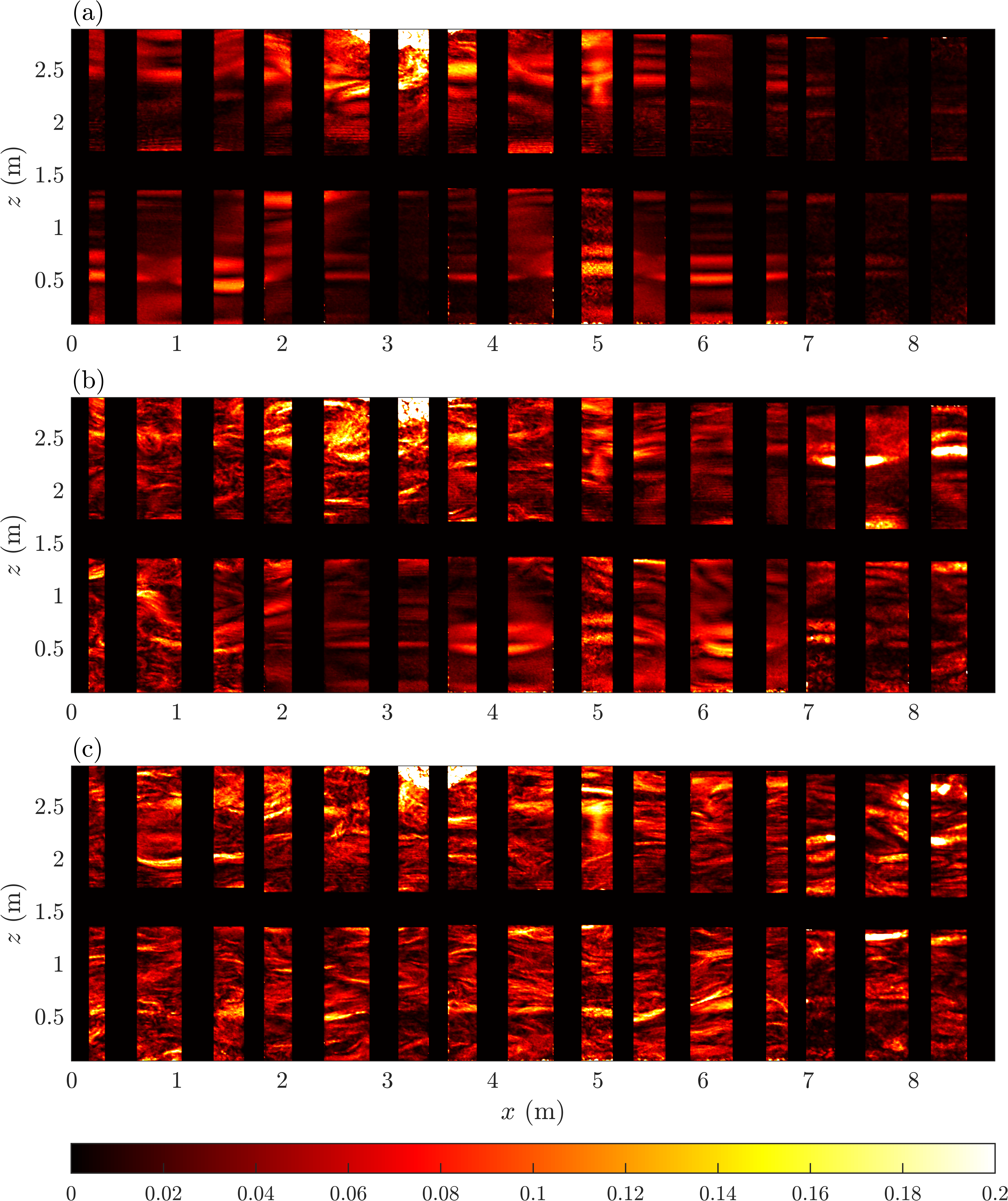}
    \caption{Magnitude of buoyancy gradient $|\nabla b|$ in s$^{-2}$ from BOS at (a) $t = 181$ s, (b) $t = 398$ s, (c) $t = 800$ s. Shading is from $0$ s$^{-2}$ (dark red) to $0.25$ s$^{-2}$ (bright yellow). Black shading represents missing data.}
    \label{fig:rawbos}
\end{figure}

We begin the results by showing $\partial b/\partial x$ from the BOS in Fig. \ref{fig:rawbos} at three different times. The first frame is shortly after the forcing begins, the second frame from a spin-up phase, and the final one being once the system has reached a sort of saturation state. A full video of the entirety of the experiment may be found in the supplemental material.

Shortly after tidal forcing starts we see oscillations along the interfaces of our thin density layers. These oscillations are largest along diagonal lines, in accordance with beam waves. Thus even though our stratification is not truly linear the waves still behave similarly to how they would in a continuously stratified fluid. The beams propagate from the topography and reflect at the boundaries in the lower left hand corner of the tank and then at the bottom on the right hand side of the tank. 

If we now turn our attention to Fig \ref{fig:rawbos}(b), which is roughly nine forcing periods after Fig \ref{fig:rawbos}(a), we see that the beam has become obfuscated behind patches of turbulence and overturning. These dynamics are concentrated at locations where the beam reflects off the walls of the tank and intersects itself. There are also turbulent patches near the topography, consistent with forcing at a large excursion number. Before long, these overturns spread throughout the tank, as shown in Fig. \ref{fig:rawbos}(c). By this point, any sort of individual wave profile is completely hidden behind the global nonlinear dynamics. Regions of overturning and mixing are no longer strictly isolated to the vicinity of the boundaries, but fill up the full domain. We consider this to be a sort of saturation state of the experiment. 
 
\subsection{Frequency Analysis}

\begin{figure}[h!]
    \centering
    \includegraphics[width=0.9\textwidth]{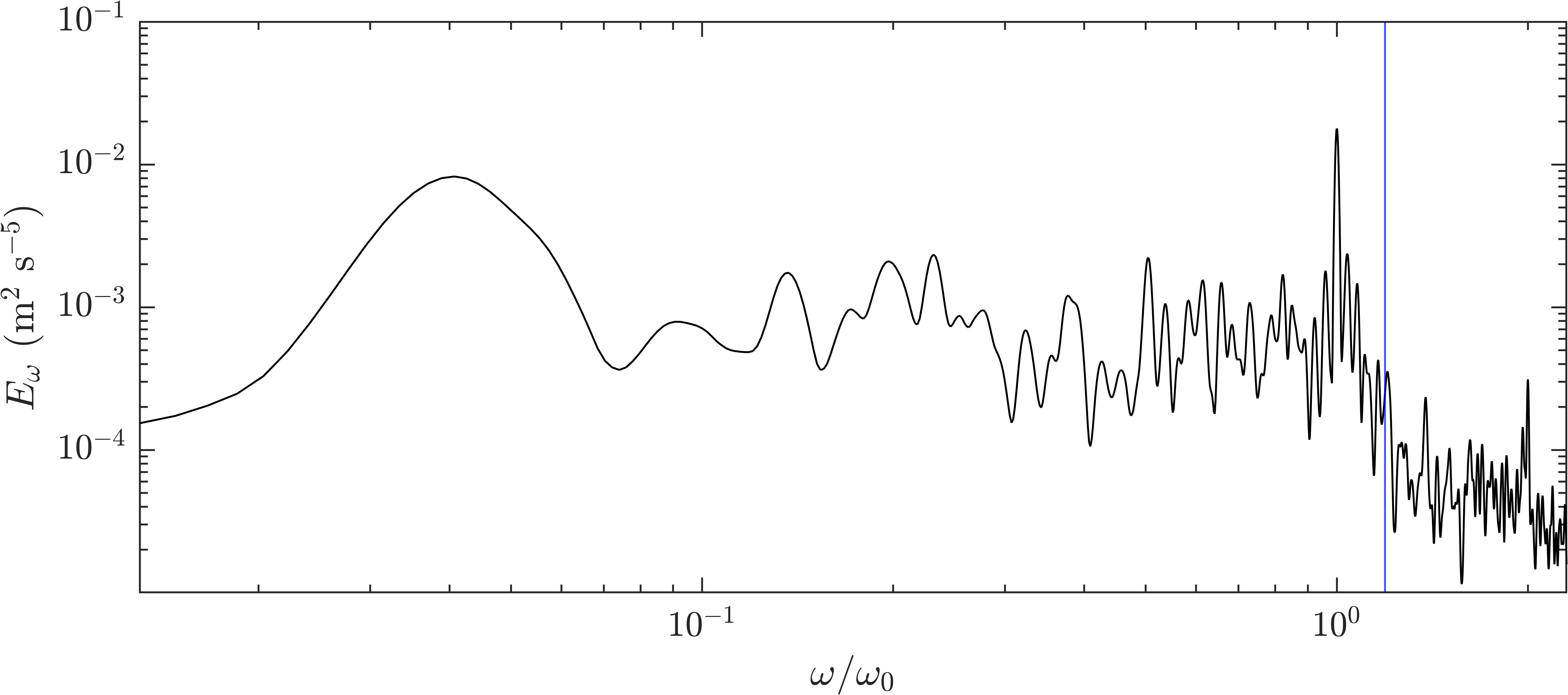}
    \caption{Power spectral density of buoyancy perturbation $b$ recorded from PME conductivity probe over the full experiment. Spectrum is calculated using Welch's method, and frequencies are normalized by the forcing frequency $\omega_0$. Vertical dashed line denotes $N$.}
    \label{fig:probespectrum}
\end{figure}

In order to gain insight into the dynamics observed in this nonlinear soup, we turn our attention to the stationary conductivity probe. As a reminder, this probe sampled continuously throughout the duration of the experiment. Using the linear fit from Fig. \ref{fig:stratification} and the depth of the probe, we can convert the density values to a buoyancy perturbation $b$. We present the energy distribution of the buoyancy perturbations through a spectrum in Fig. \ref{fig:probespectrum}. Our spectrum shows an obvious narrow peak at the forcing frequency, expected for a monochromatic forcing. Since this forcing is very close to $N$, we do not observe much of a power law in the thin range between the forcing scales and the upper limit of internal waves. Beyond $N$, the spectral values fall off quickly. However, if we turn our attention to the frequencies beneath $N$, and furthermore beneath the forcing frequency, we observe a large and roughly white spectrum. The energy is not dominated by isolated peaks, but rather is well distributed across a range frequencies. And as these frequencies fall within the admissible range set by the dispersion relation, they represent dynamics which are candidates for linear internal waves. 
 
\begin{figure}[h!]
    \centering
    \includegraphics[width=0.5\textwidth]{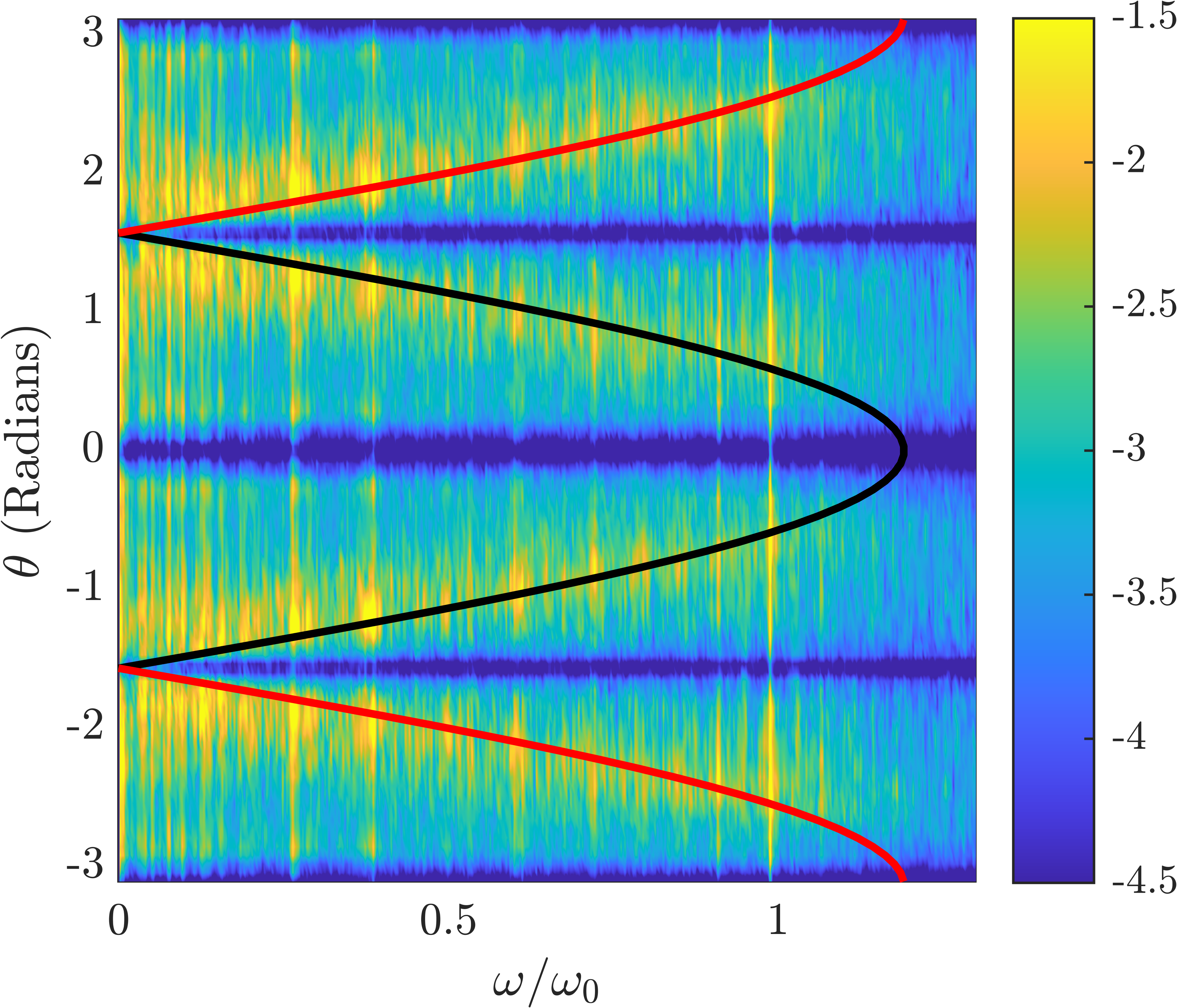}
    \caption{Power spectrum (in arbitrary units) of $\partial b/\partial z$ taken from BOS over full duration of the experiment, and averaged over horizontal angle $\theta$. Black and red curves are positive and negative roots of the dispersion curve, respectively.}
    \label{fig:disprel}
\end{figure}

Simply having these concentrations of energy at low frequencies does not guarantee that what we are observing is a result of wave turbulence, or that these low frequency motions are even waves for that matter. To check if the cascade across frequencies is through waves, we compare our full spatiotemporal dynamics to the internal wave dispersion relation. We first transform $\partial b/\partial z$ from the BOS in time and space. As we have missing spatial data, we use a nonuniform FFT (NUFFT) for the spatial transforms. The wavenumber arrays for the NUFFT's are consistent with arrays for standard FFT's. After transforming, we take the absolute square to obtain a power spectral density, and then average over thin wedges of the horizontal angle $\theta = \mathrm{arctan}(m/k)$. This gives us a spectrum as a function of $\omega$ and $\theta$. We then superimpose this spectrum with the two roots of the dispersion curve to compare the distribution of our energetic modes with the angles and frequencies predicted by linear wave theory. The result is shown in Fig. \ref{fig:disprel}. 

The most energetic subharmonic frequencies in our signal are consistent with angles predicted by the dispersion curve, in spite of the spectral leakage caused by the NUFFT's. This validates that the energetic low-frequency motions are indeed waves. Importantly, many of the same frequencies have peaks at positive and negative $\theta$, meaning both upward and downward propagation. This is significant as it suggests that our waves remain energetic after reflection, in contrast to prior experiments at a lower $Re$, where viscosity significantly damped the waves by the time they traversed the tank. Since we use reflections as a proxy for waves born at other topographies, the presence of upward and downward propagating waves indicates our wavefield approximates the ocean interior. Furthermore, the concentration of the low frequency spectral peaks along the dispersion curve is a sign of an energy cascade towards lower frequencies. Assuming this cascade is driven by wave-wave interactions, it should be possible to identify triads of frequencies engaging in the energy transfers. 

\begin{figure}[h!]
    \centering
    \includegraphics[width=0.5\textwidth]{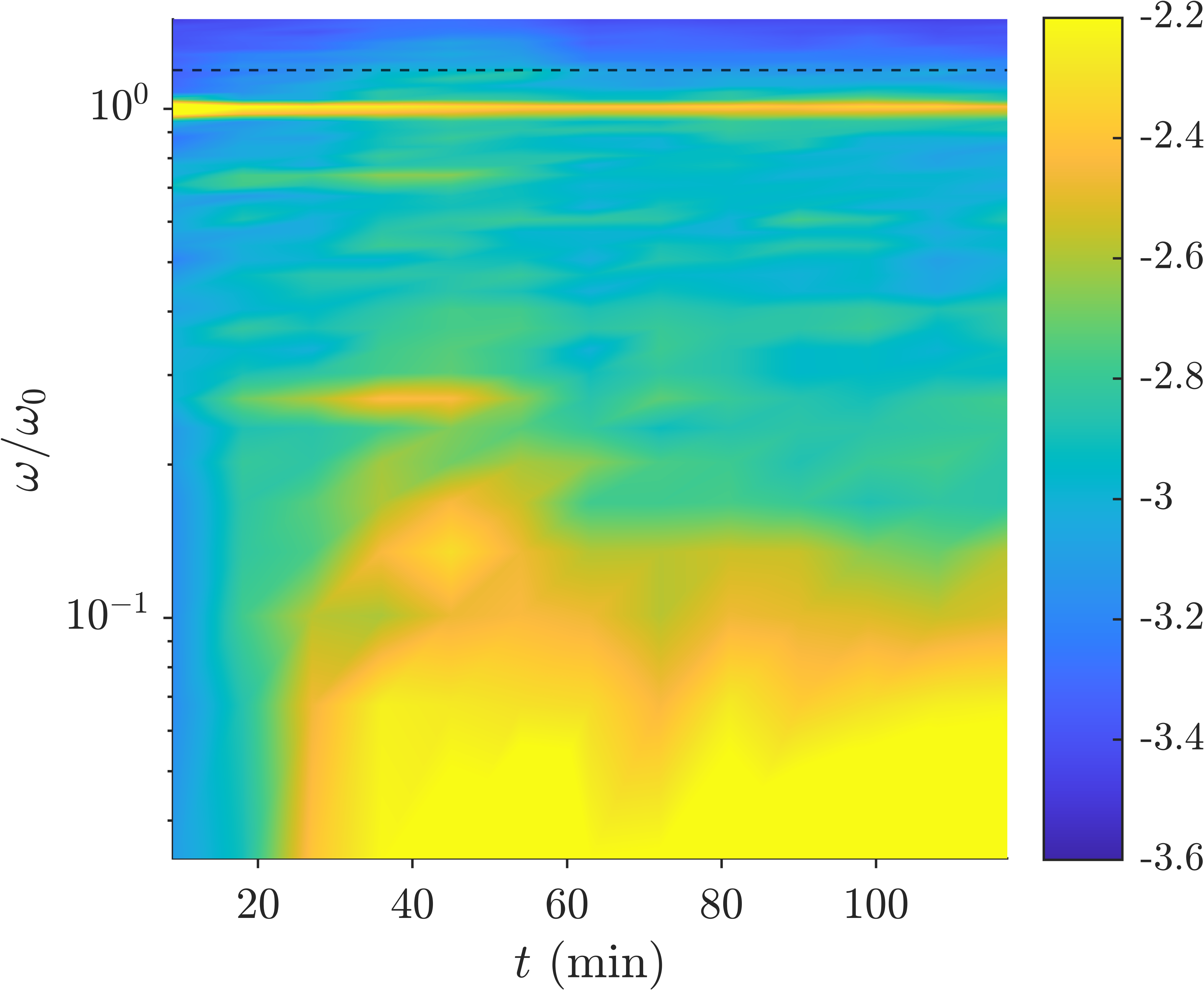}
    \caption{Spectrogram of first SPOD mode eignevalue $|\lambda_1|$ evaluated over 18 minute blocks with 50\% overlap, using both components of $\nabla b$ from the BOS. The y-axis is in a log scale. Horizontal dashed line denotes $N$.}
    \label{fig:spectrogram}
\end{figure}

To get an idea of how the different temporal scales develop within our system, we perform a spectrogram analysis on the BOS data. Since we are interested in wave-wave interactions, we would like to be able to prioritize the wave dynamics in our time-frequency analysis. We accomplish this using spectral proper orthogonal decomposition (SPOD). SPOD is an extension of standard POD, where a signal is decomposed into a sum of POD modes $\phi_n$. These modes are the solutions to an eigenvalue problem for the covariance matrix, and the corresponding eigenvalues $\lambda_n$ measure the variance of the $n$-th mode. SPOD transforms the spatiotemporal data in time via Welch's method and then performs POD for each frequency. The resulting eigenmodes and eigenvalues then capture the spatial structures that contain the variance for each frequency, with the first mode being the most dominant (which we validate for our dataset). The other benefit of SPOD is that we can input both components of $\nabla b$ from the BOS. This is similar to the use of all velocity components where the scalar product used for normalisation refers to the turbulent kinetic energy  \cite{towne2018spectral}. This also improves convergence and serves as a proxy for spatio-temporal values of $b$.
%In addition, provided that the angle of the waves spans angles in the range $\theta=[0,2\pi]$ where gradients information contained in both direction might be relevant. 
Since Fig. \ref{fig:disprel} showed us that at subharmonics the dynamics are predominantly waves, we employ SPOD to isolate these wave motions and measure the ``energy" (variance) at each frequency. Performing this algorithm over a sliding window provides with a time-frequency spectrum of the dominant spatial structures, shown in Fig. \ref{fig:spectrogram}. 

At early times, we see essentially all of the energy concentrated at the forcing frequency. There are two peaks at subharmonic frequencies of roughly $0.73\omega_0$ and $0.27\omega_0$, which satisfies the three-wave resonance condition. This interaction falls under the more general class of TRI. Other fainter peaks are visible, which can be matched to different triads. At slightly later times, the lowest daughter wave grows in amplitude, as does the energy content at even lower frequencies. Curiously, the wave at $0.23\omega_0$ decays after this, but the lower frequencies do not. Rather the energy content at low frequencies is retained throughout the duration of the experiment. This suggests that at later times there is a plethora of wave-wave interactions. Since the spectrogram changes little after the 60 minute mark, we consider this to be a nonlinear saturation state of the system, and thus we focus our attention towards the end of the experiment for the subsequent analysis. 

\subsection{Spatio-temporal Analysis}

All of the prior analysis gives us good information about the temporal behavior of the dynamics in our system, but the oceanic energy cascade is really a matter of transport across spatial scales, in particular the vertical. Additionally, three-wave interactions have a condition on the wavevectors as well as the frequencies. Therefore, in order to understand the role that wave turbulence plays in transferring energy across length scales, we need to start turning our attention towards the spatial character of our dynamics. While we employed NUFFT’s for the dispersion relation spectrum, the algorithm is not well-suited for estimating individual wavenumbers. The leakage from the NUFFT's is partially smoothed out by the angle averaging, but it one desires the specific values of wavenumbers at a fixed frequency, the leakage is too problematic. 

    \begin{figure}[h!]
    \centering
    \includegraphics[width=0.9\textwidth]{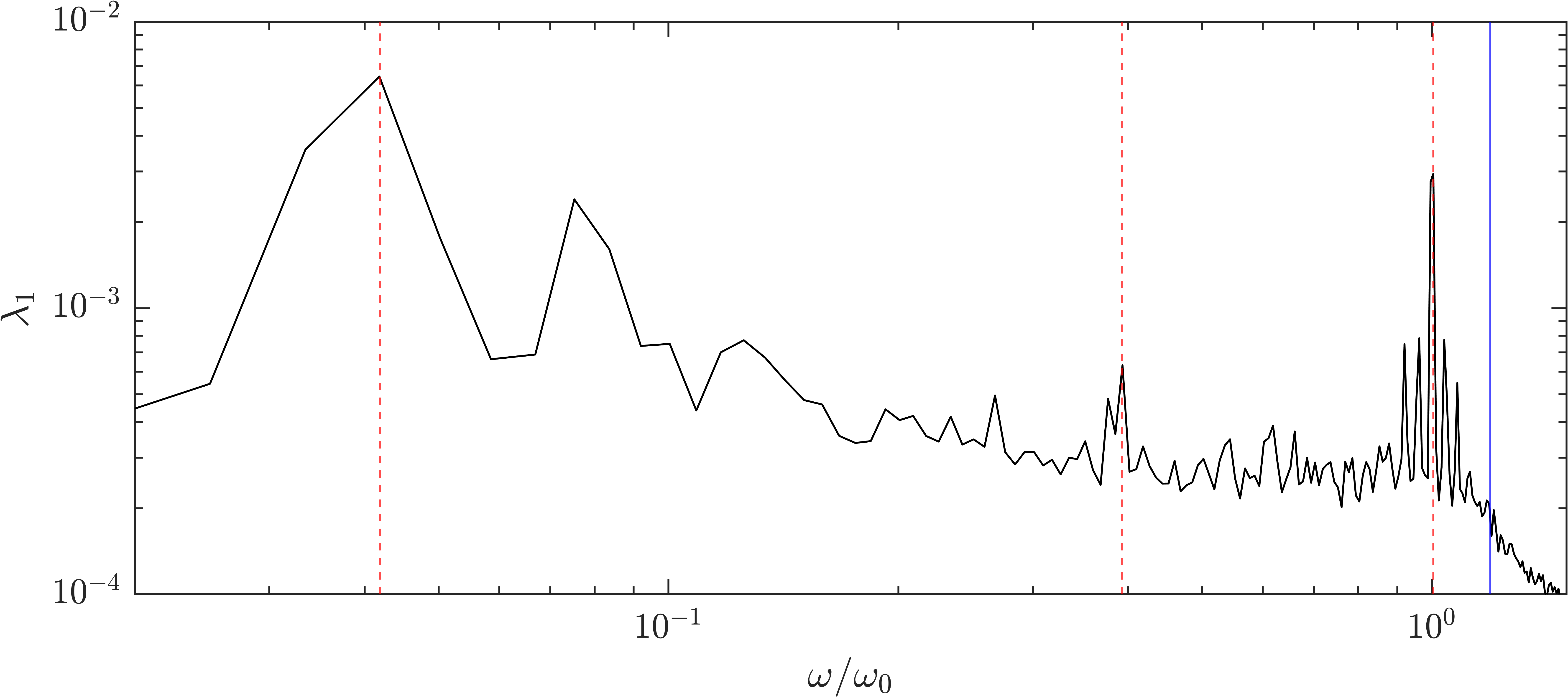}
    \caption{First SPOD mode eigenvalue $\lambda_1$ calculated on the concatenated $\nabla b$ array over the final 72.5 minutes of BOS data. SPOD is performed using two segments with 50\% overlap. Blue vertical line denotes $N$. Red vertical dashed lines denote the three frequencies from Fig. \ref{fig:spodmodes} and Fig. \ref{fig:autocorr}.}
    \label{fig:spodspectrum}
\end{figure}

We previously examined the SPOD eigenvalues to measure variance at individual frequencies, and we now consider the first SPOD eigenmode $\phi_1$. This mode contains the spatial structures that dominate the variance at a given frequency, and we can use them to determine relationships between spatial and temporal scales. To simplify the computations, we first condense the BOS arrays by excluding all missing values. It is also necessary to omit the layer of fluid near the surface at depths comparable to the topographic height, as the tidal forcing causes severe aliasing in the SPOD. We again concatenate both components of $\nabla b$, which get mapped individually to horizontal and vertical SPOD modes $\phi_{1_x}$ and $\phi_{1_z}$.
Once SPOD has been performed, we map the results back to the original domain by reintroducing the missing data to the SPOD modes, but still omit the near-surface layer. This will inherently discretize the wavenumbers in terms of a slightly smaller domain than the full tank, but as we will show this does not taint the analysis. Our analysis is performed over the final 72.5 minutes of the experiment, which we consider to be a fully developed nonlinear state based on Fig. \ref{fig:spectrogram}. 

In Fig. \ref{fig:spodspectrum} we show the SPOD eigenvalue $\lambda_1$ vs frequency. The results are consistent with the probe spectrum (Fig. \ref{fig:probespectrum}) in that the subharmonic frequencies are roughly white in energy content, with large peaks at very low frequencies. Compared to the probe spectrum, the SPOD spectrum favors the peaks adjacent to the forcing frequency. These peaks correspond to frequencies offset from the forcing frequency by amounts equal to the frequencies of the two lowest peaks ($0.042\omega_0$ and $0.084\omega_0$), which we will discuss later.

\begin{figure}[h!]
    \centering
    \includegraphics[width=0.9\textwidth]{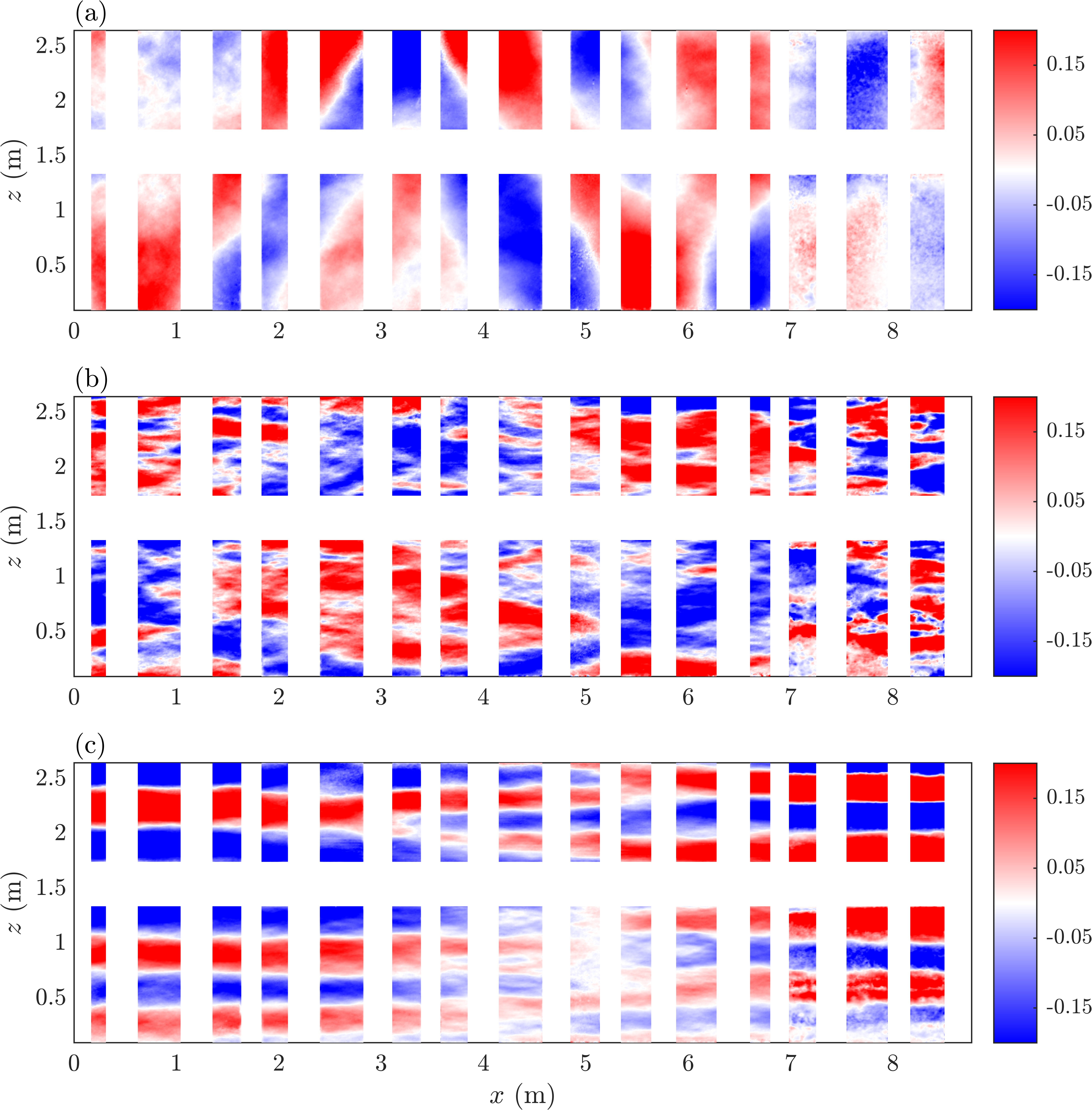}
    \caption{Real part of first SPOD mode calculated on the concatenated $\nabla b$ array over the final 72.5 minutes of BOS data at (a) $\omega = 0.042\omega_0$, (b) $\omega = 0.393\omega_0$, (c) $\omega = \omega_0$. SPOD is performed using two segments with 50\% overlap. In (a) we show $\phi_{1_x}$, in (b) and (c) we show $\phi_{1_z}$.}
    \label{fig:spodmodes}
\end{figure}

For the purpose of the SPOD analysis it is convenient to identify several frequencies of interest to examine in detail. We choose three frequencies represented by red vertical lines in Fig. \ref{fig:spodspectrum}. The lowest, $0.042\omega_0$, appears in all of our spectral analysis as a large peak. The second, at $0.393\omega_0$, is an isolated peak in the ``continuum" of frequencies. The highest is the forcing frequency $\omega_0$. In Fig. \ref{fig:spodmodes}, we show the first SPOD modes at these frequencies. 

The spatial structure of the forcing frequency reveals a very coherent beam wave pattern which is strongest near the topography and propagates down. This beam pattern illustrates how waves at our forcing scales treat the domain as if it were continuouslt stratified, as there is no evidence of the individual layers in Fig. \ref{fig:spodmodes}(a). The beam reflects at locations consistent with the raw BOS (Fig. \ref{fig:rawbos}) and regions of strong overturning, namely the lower left corner of the tank. Near the right wall, the beam becomes somewhat less coherent after reflection, which may be partially due to viscosity or interference from other frequencies. From the dispersion relation, it stands to reason that the phase velocity is up and to the right on the right flank, and up and to the left on the left flank. It is also evident that the vertical length-scale of the wave is comparable to the depth of the tank. 

The second panel shows the SPOD structure for the isolated peak at $0.393\omega_0$. This profile is far less coherent that at $\omega_0$. There are some elements of a linear wave pattern as there are obvious diagonal bands, but these are blurred by smaller scale non-wave motions. This may be in part due to the oscillations at the interfaces, or overturning. Finally the lower panel shows the SPOD mode at the low frequency $0.042\omega_0$. This spatial structure appears at first glance to be composed of thin horizontal layers. But on closer inspection one can see that the mode changes sign between the left and right walls of the tank at a fixed depth (most easily seen directly above the horizontal missing data, at roughly $z = 1.8$ m). This pattern is consistent with linear internal waves at very low frequencies, where the beam is nearly flat. The peak values being adjacent to the walls suggest this wave is described in the horizontal by a cosine, and has a horizontal wavelength of roughly twice the tank length. In other words, this is the gravest mode in the horizontal. Furthermore, the thickness of these beams is sufficiently larger than the density layers to rule out strictly inferfacial motions. Of note is the region near the middle of the tank where the SPOD mode changes sign, particularly within $x = 4-6$ m, and below the horizontal metal bar. This region appears to be slightly slanted, and its angle and location are consistent with the right beam of the wave at $\omega_0$. This hints at an area of intersection and possible elevated shear between the two modes. Since the SPOD spectrum highlighted peaks at $\omega_0-0.042\omega_0$ and $\omega_0+0.042\omega_0$, we can interpret those modes as being born from the interactions, which may be concentrated along the intersection.  
 
\begin{figure}[h!]
    \centering
    \includegraphics[width=0.9\textwidth]{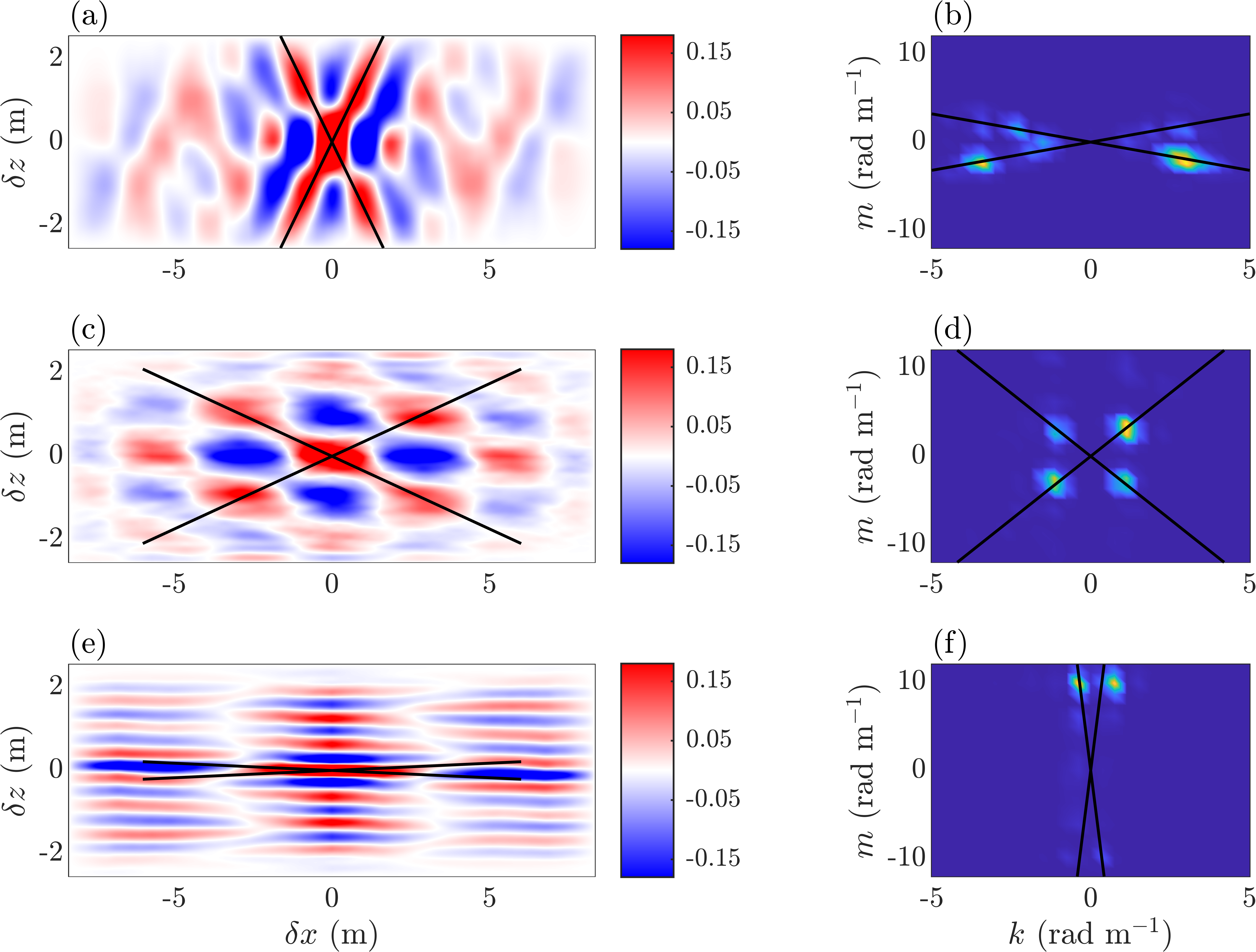}
    \caption{(a,c,e) Magnitude of autocorrelation from SPOD modes in Fig. \ref{fig:spodmodes}, respectively. Black crosses represent phase line angles predicted by the dispersion relation for each frequency. $\delta x$ and $\delta z$ denote the horizontal and spatial separation, respectively. (b,d,f) 2D spectral density (in arbitrary units) of panels (a,c,e), respectively. Black crosses represent ratio of $m$ and $k$ predicted by the dispersion relation for each frequency. Peak values are shaded yellow.}
    \label{fig:autocorr}
\end{figure}

From Fig. \ref{fig:spodmodes} we can conclude that the SPOD modes reveal the wave pattern in some of the frequencies, such as the forcing frequency, but for other frequencies the spatial structure is still obfuscated amidst non-wave dynamics. We need to further filter the results to truly isolate the wave profile at a given frequency. To do that, we take the autocorrelation of the SPOD modes. Autocorrelations have been used in prior experimental work to isolate wave dynamics \cite{rodda2022experimental, savaro2020generation}. We first fill in missing data using a discrete cosine transform interpolation \cite{garcia2010robust,wang2012three}. We feel justified in interpolating the SPOD modes since they identify dominant spatial structures, and we are not attempting to interpolate small scale turbulence. Once done, we calculate the autocorrelation for each SPOD mode at each frequency. In accordance with the dispersion relation, the waves should be most correlated with themselves along phase lines, with corresponding wavenumbers. In the left column of Fig. \ref{fig:autocorr} we show the magnitude of the autocorrelation of the three aforementioned frequencies. To confirm consistency with linear wave theory, we superimpose the group velocity angle from the dispersion curve for each frequency.

The autocorrelation at the forcing frequency maintains a beam profile, with angles in excellent agreement with the dispersion relation. In contrast, the autocorrelation at $0.393\omega_0$ follows a checkerboard pattern and lacks any obvious beam structure. This pattern is expected for modal waves, internal waves with wavenumbers quantized by the size of the finite domain. In principle the waves at $\omega_0$ should also be modal, but the autocorrelation shows a beam with some checkerboard features. This may be attributed to the primary beam losing energy to wave-wave interactions before it can fully reflect and form a standing wave pattern. Returning to $0.393\omega_0$, the pattern does obey the dispersion curve. Finally, the lowest frequency, $\omega = 0.042\omega_0$, also displays a checkboard pattern but with a greater disparity between vertical and horizontal length scales. This is expected by the dispersion relation, and while difficult to see, the pattern has the expected angles. In all cases, we see the autocorrelation provides sufficient filtering to isolate the wave profile.

We have now set the stage to extract spatial scales from our temporal scales. With the autocorrelation capturing the wave structure, we can Fourier Transform the autocorrelation in space and identify the peak wavenumbers. We again wish to maintain veracity with linear wave theory, and so we show the FFT of the autocorrelation in the right column of Fig. \ref{fig:autocorr}. We compare against the predicted ratio $m/k$ to ensure that our wavenumbers agree with the dispersion relation. And as seen, in all cases the peak wavenumbers lie on or near the dispersion curve. This demonstrates our algorithm is reliable for capturing the wave spatial scales. Note that the spectrum is in arbitrary units, since we are not interested in the actual spectral amplitudes but rather the peak wavenumbers.

We are now in position to extract spatial scales from all SPOD modes up to the BV frequency, however since we perform SPOD on both components of $\nabla b$, we need a method to decide which component to retain for a given frequency. We address this by performing the autocorrelation and FFT on each component of $\nabla b$ individually, and save the wavenumbers. To choose whether to use $(k,m)$ from $\phi_{1_x}$ or $\phi_{1_z}$, we plug the measured wavenumbers into the dispersion relation to get a calculated value of $\omega$ for each component. We retain the component with the smallest error between the calculated frequency and the frequency from the SPOD array. We apply this algorithm to every frequency up to $N$, and obtain an array of wavenumbers for each frequency. Note that since Fig. \ref{fig:disprel} revealed the same frequency peaks at multiple angles, we consider all sign combinations for the measured $(k,m)$, as we expect the waves to engage in reflections and propagate in multiple directions.  

\begin{figure}[h!]
    \centering
    \includegraphics[width=0.9\textwidth]{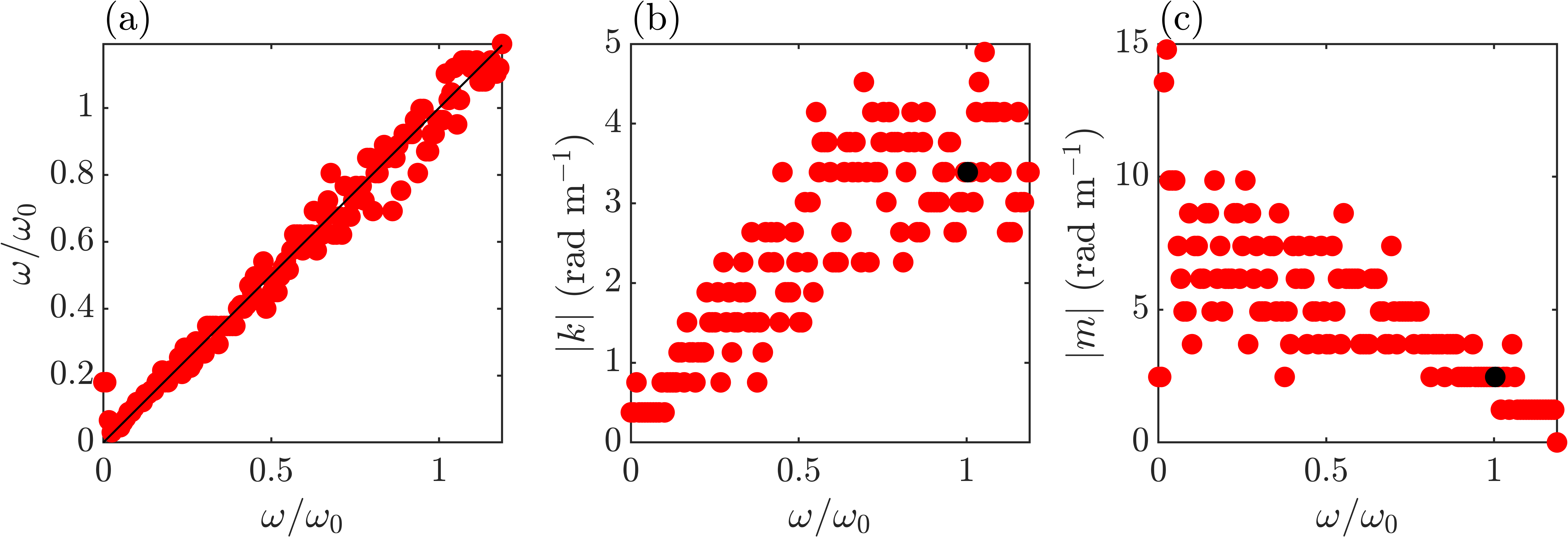}
    \caption{(a) SPOD frequency vs calculated frequency from autocorrelation FFT wavenumbers. Input frequency from SPOD array is on the x-axis, calculated frequency is on the y-axis. Black line denotes one to one relationship. (b) Magnitude of horizontal wavenumber from autocorrelation FFT vs SPOD array frequency. Black dot denotes the forcing frequency $\omega_0$. (c) Magnitude of vertical wavenumber from autocorrelation FFT vs SPOD array frequency. Black dot denotes the forcing frequency $\omega_0$.}
    \label{fig:spatiotemporal}
\end{figure}

Before comparing spatial and temporal scales, we do one final consistency check with linear wave theory. We take the wavenumbers we have calculated and plug them back into the dispersion relation, and compare the output frequency to the frequency from the SPOD array. This is shown in Fig. \ref{fig:spatiotemporal}(a). As seen, we have excellent agreement between our wavenumbers and linear wave theory. The only main discrepancies are a few very low frequencies (lower than those plotted in any of the spectra), and several frequencies just beneath the forcing. The overall agreement though speaks to the reliability of our approach, and the energetic regime from Fig. \ref{fig:spectrogram} is very well captured. Thus we feel confident drawing conclusions on the wavefield from our calculated wavenumbers. 

Now that we have reason to believe that these spatial scales are consistent with wave dynamics we can compare the spatial to the temporal scales of the waves. In Fig. \ref{fig:spatiotemporal}(b) and (c) we plot the wave frequency versus the magnitude of the horizontal and vertical wavenumbers, respectively. The overall relationship between $|k|$ and $\omega$ is that as we move to smaller frequencies, we move to smaller horizontal wavenumbers. This tells us that the subharmonics we create are at larger horizontal scales, consistent with Fig. \ref{fig:spodmodes}(c). At frequencies close to and above the forcing frequency, the relationship between $\omega$ and $|k|$ is less clear. We observe both longer and shorter waves in both directions. 

But as aforementioned, it is the vertical wavenumber that tells the story of energy transfer in the ocean. From Fig. \ref{fig:spatiotemporal}(c) we see the opposite relationship as the prior figure. As $\omega$ decreases, $|m|$ grows larger, meaning the subharmonics are at smaller vertical scales. The transfer spans just shy of a decade in $|m|$, with some larger values of $|m|$ at the lowest frequencies that diverged from the dispersion curve in panel (a). This tells us is that by producing subharmonics we are moving energy to smaller vertical scales, and corroborates the idea of wave wave interactions moving energy to smaller spatial scales. Therefore, even without a forward cascade in $\omega$, wave turbulence can still produce a downscale transfer in $m$. 

\subsection{Three-Wave Interactions}

\begin{figure}[h!]
    \centering
    \includegraphics[width=0.5\textwidth]{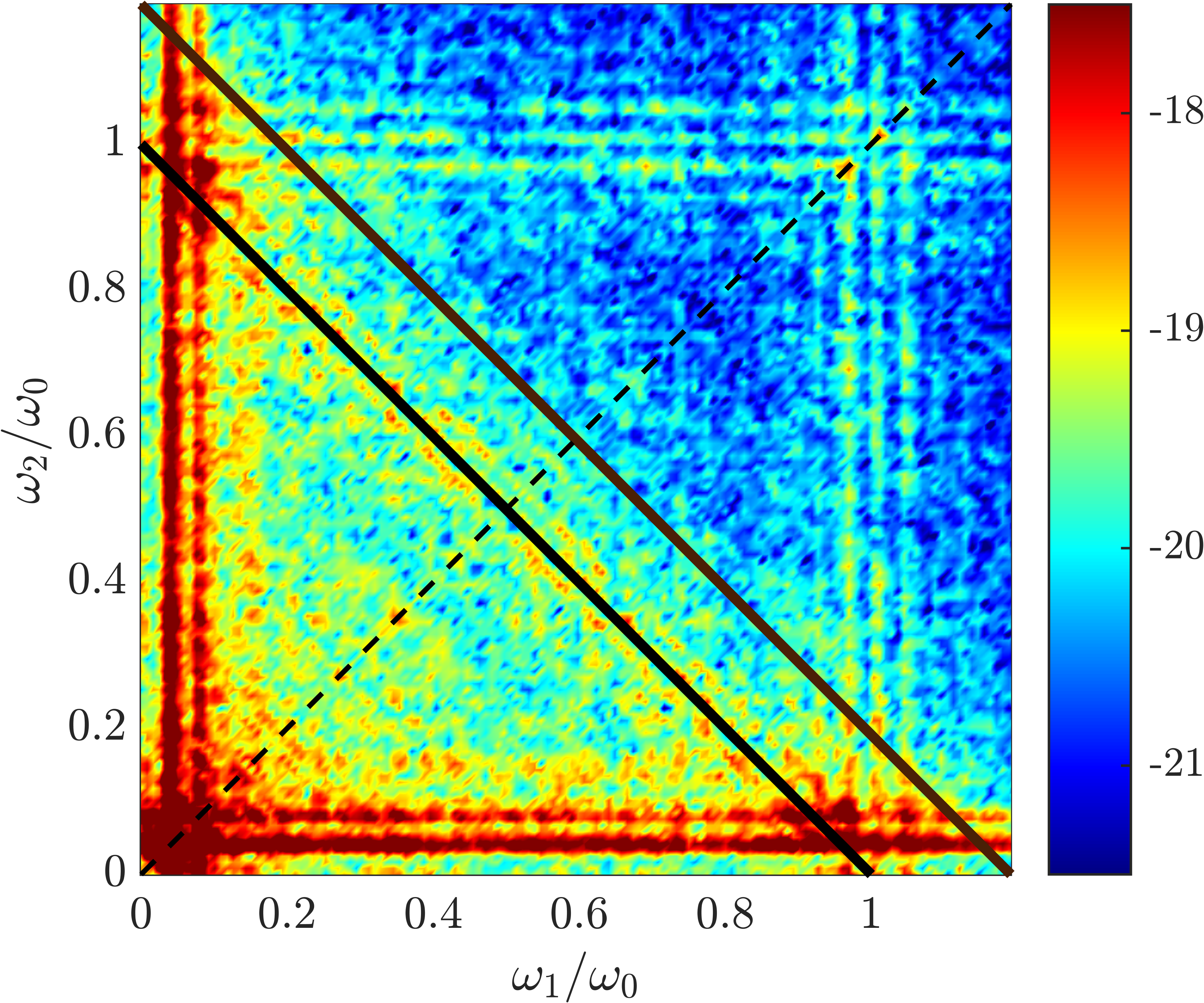}
    \caption{Magnitude of mode bispectrum $|\lambda_1|$ calculated using BMD on the final 72.5 minutes of BOS data from camera 1. Plot is symmetric across the diagonal dashed line $\omega_1 = \omega_2$. Solid black line denotes triads with $\omega_3 = \omega_0$. Solid purple line denotes triads with $\omega_3 = N$.}
    \label{fig:bmd}
\end{figure}

With the frequencies and wavevectors identified in the prior section, we have the tools to start investigating how the new waves form. Before searching for viable triads based on our spatiotemporal scales, it is practical to first get an idea of the most meaningful interactions in our system. If we assume three-wave resonances it should be possible to identify correlations between the waves in a given triad. This is not possible through a simple FFT analysis but it is possible using higher order spectral methods. In particular the bispectrum can measure phase correlation between three Fourier modes whose frequencies obey the triad resonance condition. Computing a bispectrum is typically done by Fourier transforming in time and then taking spatial averages; however, we elect to use a slightly different algorithm called bispectral mode decomposition (BMD). This process calculates the bispectrum using spatial correlations, in the spirit of a higher order spectral SPOD. In doing so it determines a bispectral density, which enables computation of the bispectrum over the entire field of view. Given the computational difficulty of performing a BMD analysis we only perform BMD on a single camera rather than the entire field of view. However, we have verified that the results are consistent across all cameras and therefore we choose camera one, the leftmost camera, to be the representative of the BMD analysis. This camera films the location where the wave at $\omega_0$ reflects off the corner of the tank. Our results are shown in Fig. \ref{fig:bmd}.

In the bispectral plot the x and y-axes represent the individual frequencies $\omega_1$ and $\omega_2$ in a triad, normalized by the forcing frequency. The third frequency in the triad, $\omega_3$, is given as the sum of the x and y-coordinates at any location in the plot, while the color shading represents a measure of the correlation between the three Fourier modes. If we assume that we have relationships of the form $\omega_1 + \omega_2 = \omega_3$, one would expect to see correlation between these waves as one side of the equal sign serves as an energy source for the other side. We plot all triads where $\omega_3 = \omega_0$ as the lower solid black line, and all triads where $\omega_3 = N$ as the upper solid purple line. Note that the plot is symmetric across the line $\omega_1 = \omega_2$.
 
From the bispectrum we observe that the largest phase correlation is for triads where $\omega_3$ is less than or equal to $N$, which is consistent with interactions between linear internal waves. For $\omega_3 > N$, the correlation falls off by at least an order of magnitude. There are isolated superbuoyant peaks, such as at $2\omega_0$. These may represent bound waves (sometimes referred to as evanescent waves) born from nonresonant interactions in which $\omega_1 + \omega_2 = \omega_3$, but $\omega_3$ does not satisfy the dispersion relation. They may also be an artifact of aliasing in the BMD. Returning to linear wave triads, there are elevated spots of correlation with triads involving the forcing frequency, as well as some isolated large values of correlation at frequencies beneath the forcing. The overall trend is large correlation across a range of scales. Most notably however, observe that the largest correlation involves two frequencies at $0.084\omega_0$ and $0.042\omega_0$. As a reminder, these frequencies presented as peaks in Fig. \ref{fig:spodspectrum}, and we showed the SPOD structure of $0.042\omega_0$ in Fig. \ref{fig:spodmodes} and Fig. \ref{fig:autocorr}. These frequencies have large correlation with roughly every frequency up to and slightly beyond $N$. Unlike the superbuoyant motions, we do not believe that these bands of elevated correlation are an artifact of aliasing, which we will explain below. This large and distributed correlation hints at a sort of strong background triad interaction involving extremely low frequencies.
 
We now shift our attention towards identifying the types of triads within our system. Equipped with frequencies and wavenumbers from the SPOD, we iterate through all frequencies that satisfy the temporal and spatial resonance conditions simultaneously. To accommodate experimental uncertainties, we set relative error tolerance on the temporal and spatial resonance conditions of 3\% and 12\% respectively. These values are chosen based on a middle ground between overly strict conditions omitting meaningful triads, and overly lenient conditions admitting non-meaningful interactions. Our definition of meaningful and non-meaningful is based on consistency between the triads we calculate, and the results from the BMD. 

Once we have identified all the triads in our system, we can start categorizing them based off of nonlocal and local interactions. We do this following the methodology of Wu and Pan \cite{wu2023energy}, which we restate here. We first take a resonant triad and order the frequencies and vertical wavenumber (magnitude) from high to low as $(\omega_\mathrm{L}, \omega_\mathrm{M}, \omega_\mathrm{H})$ and $(|m|_\mathrm{L}, |m|_\mathrm{M}, |m|_\mathrm{H})$. We then set criteria for the three nonlocal interactions based on reference values. 

\begin{description}
    \item[PSI:] $|m_\mathrm{M}|/|m_\mathrm{L}| > 2$ and $0.5 \leq \omega_\mathrm{M}/\omega_\mathrm{H}<0.55$
    \item[ES:] $\omega_\mathrm{M}/\omega_\mathrm{L} > 2$ and $0.5 \leq |m_\mathrm{M}|/|m_\mathrm{H}|<0.55$
    \item[ID:] $\omega_\mathrm{M}/\omega_\mathrm{L} > 2$ and $|m_\mathrm{M}|/|m_\mathrm{L}|>2$
\end{description}

Any triad which does not meet any of the above criteria would be classified as local. There are additional, but subtle conditions baked in to the notion of the nonlocal interactions. For example, as shown in Fig. \ref{fig:nonlocaltriads}, in induced diffusion the lowest frequency wave has the smallest vertical wavenumber (in magnitude). This means that $\omega_\mathrm{L}$ and $|m_\mathrm{L}|$ must be from the same wave. The opposite is true of PSI, where $\omega_\mathrm{H}$ and $|m_\mathrm{L}|$ must be from the same wave. We include these additional constraints into our selection criteria. 

As explained in \cite{wu2023energy}, the reference values of 2 and 0.55 are chosen with the intent of being as unaccommodating to local interactions as possible. In other words as shown in Fig. \ref{fig:nonlocaltriads}, the example of ID involves a low frequency wave with $|m_\mathrm{M}|/|m_\mathrm{L}|$ much larger than 2. However we accept a triad as ID with only the mild cutoff of a scale separation of 2. Thus weakly ``nonlocal" triads would be classified as local under a selection criteria more consistent with Fig. \ref{fig:nonlocaltriads}. So if we see considerable influence of local triads under these oppressive conditions, their impact would be even greater under more realistic criteria. 

\begin{figure}[h!]
    \centering
    \includegraphics[width=0.9\textwidth]{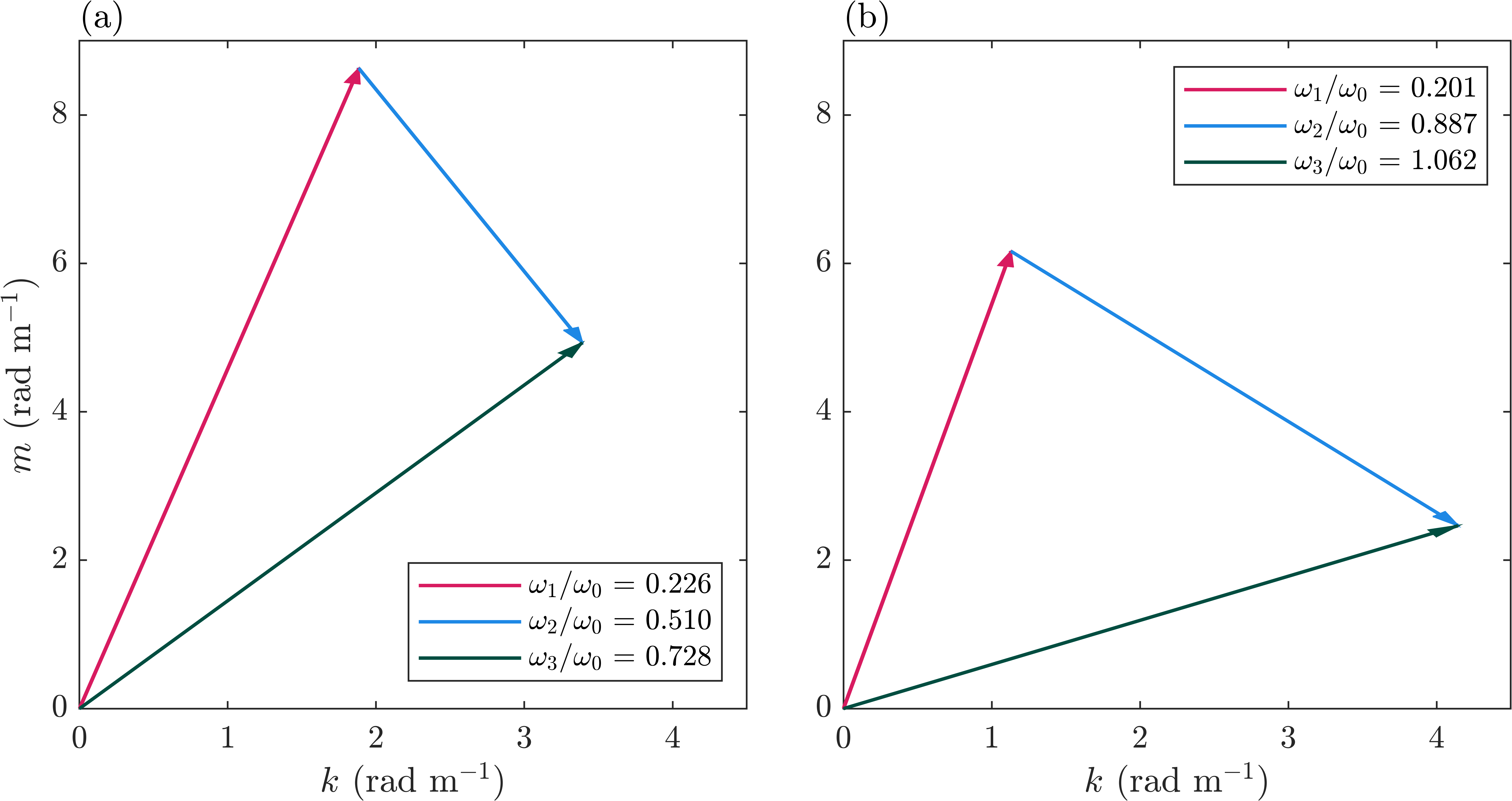}
    \caption{Magnitude of mode bispectrum $|\lambda_1|$ calculated using BMD on the final 72.5 minutes of BOS data from camera 1. Plot is symmetric across the diagonal dashed line $\omega_1 = \omega_2$. Solid black line denotes triads with $\omega_3 = \omega_0$. Solid purple line denotes triads with $\omega_3 = N$.}
    \label{fig:local}
\end{figure}

Case in point, we find the vast majority (roughly 76\%) of our triads to be classified as local. This holds true even under harsher error tolerance for the resonance conditions. Two examples of local triads in our system are shown in Fig. \ref{fig:local}. The first triad, in panel (a), represents the many triads within the ``interior" of the BMD plot. That is, the triads that fall inside the triangle formed by the line $\omega_1=\omega_2$, $\omega_3 = \omega_0$, and the dark bands of the two low frequencies. This can be seen as the continuum of modes below the forcing frequency, and tells us that much of the middle frequencies interact with each other through local interactions. As these interactions are the most plentiful, we find them to be a likely suspect for the underlying energy cascade. As shown in the second panel, (b), local interactions are also responsible for the formation of much of the energy in frequencies $\omega_0<\omega_3\leq N$. While we have only a narrow range between our forcing and BV frequency, the frequencies that do exist are likely born through local interactions. Prior work has suggested local interactions to be largely responsible for the direct cascade of energy in frequency \cite{wu2023energy}. While we can only offer some corroboration based on our thin range of frequencies between $\omega_0$ and $N$, the results we do have support the idea of local triads pushing energy to higher frequencies. 

\begin{figure}[h!]
    \centering
    \includegraphics[width=0.9\textwidth]{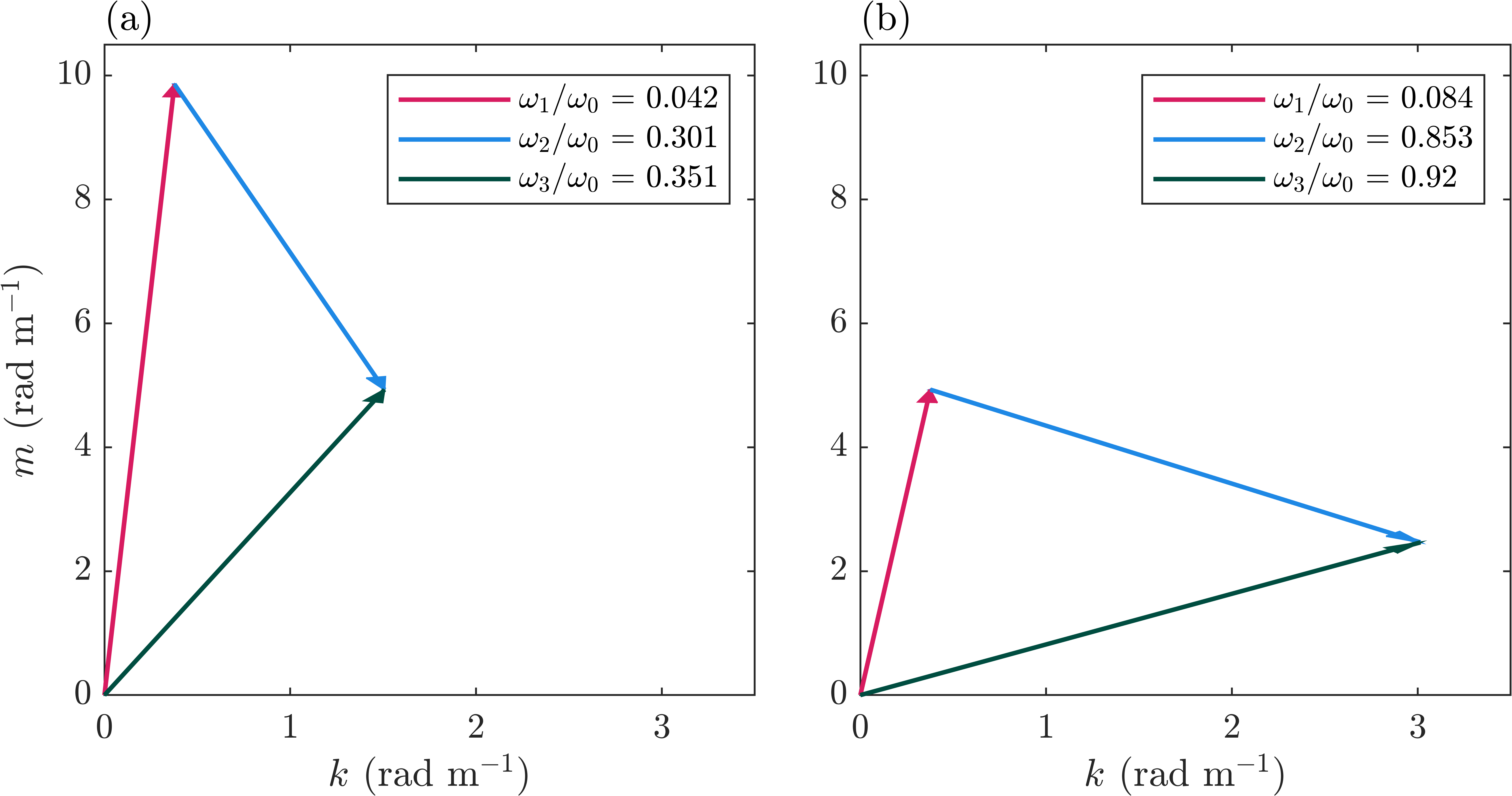}
    \caption{Magnitude of mode bispectrum $|\lambda_1|$ calculated using BMD on the final 72.5 minutes of BOS data from camera 1. Plot is symmetric across the diagonal dashed line $\omega_1 = \omega_2$. Solid black line denotes triads with $\omega_3 = \omega_0$. Solid purple line denotes triads with $\omega_3 = N$.}
    \label{fig:es}
\end{figure}

Let us now focus on the two extreme bands of low frequency correlation from Fig. \ref{fig:bmd}. As a reminder, those waves are at $0.042\omega_0$ and $0.084\omega_0$. When we apply our classification scheme to these interactions, we find that almost all of them meet the conditions for elastic scattering. An example ES triad from each is shown in Fig. \ref{fig:es}. This is consistent with the idea of elastic scattering involving a low frequency wave with a larger vertical wavenumber, which we see in Fig. \ref{fig:spodmodes} and \ref{fig:autocorr} for $0.042\omega_0$. Importantly, the number of elastic scattering triads we identify are consistent with the omnipotence of correlation these waves have in the BMD. In other words, we do not believe that the bands of elevated correlation at $0.042\omega_0$ and $0.084\omega_0$ are an artifact of aliasing, but rather represent the number of ES triads in our wavefield. And we have further consistency with Fig. \ref{fig:spodspectrum}, which shows large energy (variance) at these two frequencies. We therefore suspect them to offer considerable energy transfer in their interactions. We find elastic scattering to be the second most plentiful interaction in our system. In fact, ES accounts for nearly the entire 24\% of our triads that are nonlocal. We do have some ES triads which do not involve $0.042\omega_0$ or $0.084\omega_0$, but these two frequencies are featured the most.

But what of ID and PSI? We will start with the former, and observed very little of it. The paucity of ID can be understand based on its somewhat mutual exclusivity with ES. ES requires $\omega_\mathrm{L}$ to correspond to $|m_\mathrm{H}|$, while ID requires the opposite, $\omega_\mathrm{L}$ corresponds to $|m_\mathrm{L}|$. If we return to Fig. \ref{fig:spatiotemporal}(c), as $\omega$ gets smaller, $|m|$ gets larger. This makes it difficult to simultaneously have a small $\omega$ and a small $|m|$. The two subharmonic peaks at $0.042\omega_0$ and $0.084\omega_0$ are ill-suited for ID, as they both have large vertical wavenumbers. In general, as $\omega$ gets smaller the ratio $|m/k|$ must get larger. But at a certain point for a finite-size system $|k|$ can get no smaller, and a further reduction in $\omega$ requires an increase in $|m|$. This is true for our setup, as the smallest frequencies have the smallest horizontal wavenumbers. With that in mind, ID seems more well-suited to a system with a much larger separation between horizontal and vertical extent in which simultaneously having a large $|m/k|$ and a small $|m|$ is feasible. As an extension of this, ID may be better suited to a system with more freedom to cascade at frequencies above the forcing, where the scales are further separated from the size of the domain, and a large $|m/k|$ can include a small $|m|$. 

As for PSI, we know from the early times of Fig. \ref{fig:spectrogram} that we have a triad $0.27\omega_0+0.73\omega_0 = \omega_0$. This triad does not meet the conditions for PSI, but falls under the more general class of subharmonic generation termed TRI. This is a consequence of the MB classification neglecting viscous effects, and assuming instabilities to spawn daughter waves near half of the driving frequency. For our system a more realistic PSI (or rather, TRI) condition would require pushing the frequency bounds out further than $(0.5\omega_0, 0.55\omega_0)$. Ideally one could determine the new subharmonic frequencies with peak growth rate based on the viscosity of the fluid, and the streamfunction of the wave at $\omega_0$ \cite{joubaud2012experimental}. Unfortunately we cannot determine streamfunction from only $\nabla b$, but this is somewhat of a moot point. We find that even by broadening our PSI frequency criteria to accommodate the observed $0.27\omega_0+0.73\omega_0 = \omega_0$ does not significantly increase the contribution of PSI/TRI at late times in our system. 

Nevertheless an examination at earlier times (0-40 minutes) reveals a greater contribution of TRI, in which we relax the PSI classification scheme to admit daughter waves at frequencies between 0.2 and 0.8 times the parent wave. This includes the aforementioned $0.27\omega_0+0.73\omega_0 = \omega_0$, but we also identified two TRI triads where the mode at $0.27\omega_0$ is presumed to be the parent wave. Two of the resulting four daughter waves are similar to the extremely low frequency modes from the elastic scattering ($0.042\omega_0$,$0.084\omega_0$). If we recall Fig. \ref{fig:spectrogram}, the elevation of energy at the lowest frequencies occurred at roughly the same time that the mode at $0.27\omega_0$ attained maximum energy. At that point, it may become unstable and trigger its own TRI, which leads to the two extremely low frequency waves at $0.042\omega_0$ and $0.084\omega_0$. After that, these two new waves take over much of the triad interactions through elastic scattering. This picture points to PSI/TRI as being meaningful early in the experiment to fill the domain with enough modes to start enabling other three-wave interactions, consistent with prior experimental work \cite{rodda2022experimental, lanchon2023internal, savaro2020generation}. After that, PSI seems less relevant, as it had almost no role in the late stage interactions. As an aside, note that local triads and elastic scattering are not absent in early times, but the waves involved are less energetic.

It is worth mentioning that there is some ambiguity here in classification. By extending the realm of validity of PSI to such distant daughter wave frequencies, we are redefining interactions that would be otherwise termed local. In this sense, one could interpret our TRI as really being local interactions that behave similarly to PSI, and may in fact be born through instabilities. Alternatively, one could interpret the classification for PSI from \cite{wu2023energy} as ill-suited to a laboratory-size domain based on the role of viscous effects. This debate does not affect whether or not the interactions exist, but rather how we categorize them. The process of spawning daughter waves early in the experiment is observed whether we term it PSI, TRI, or local triads. 

\section{Discussion and Conclusions}

In our study, we sought to determine whether wave turbulence could serve as an effective first stage of energy transfer in the ocean. Assuming just tidal forcing, we found strong evidence of an energy cascade in vertical wavenumbers. This is particularly promising given the experimental restrictions of our setup. Even without having a large separation between $\omega_0$ and $N$, the preclusion of a direct cascade in frequency did not inhibit a forward cascade in wavenumber. One would expect then that with greater separation between $\omega_0$ and $N$, wave turbulence would have even more freedom to cascade energy. Nevertheless, our work speaks to the wave dynamics that can occur at frequencies below the tidal forcing. This frequency range is typically thought to be almost exclusively PSI, but at latitudes close to the equator, prior work has shown spectra with energy content spread across much more than just PSI peaks \cite{liao2012current}. Our study finds similar spectral results, and points towards underlying mechanisms on energy transfer which could be at play in the ocean.

We obtained our WT cascade starting only from a monchromatic tidal forcing. This point warrants some conversation, since it is a key distinction between our work and prior experimental studies \cite{rodda2022experimental, rodda2023internal, savaro2020generation}. Our energy injection is through an internal tidal beam, consistent with real-world observations \cite{whalen2020internal}. This appeals not only to a more applicable methodology, but also appeases the research regarding nonlinearities caused by finite-width beams \cite{brouzet2016internal}. Indeed, panel (a) of both Fig. \ref{fig:spodmodes} and Fig. \ref{fig:autocorr} showed that the wave at forcing frequency $\omega_0$ is not an individual modal wave, but rather a composition of several different modes producing a tidal beam. This can be attributed to our internal tidal forcing exciting waves at a multitude of spatial scales, as expected for internal tides in the ocean \cite{staquet2002internal}. And as seen in Fig. \ref{fig:autocorr}(b), there are several peaks in the FFT, all along the dispersion relation and all at unique values of $(|k|,|m|)$. That we attained WT from a tidal forcing speaks volumes towards its applicability in the ocean energy transfer, particularly since tidal forcing accounts for roughly half of the necessary energy input \cite{mackinnon2013diapycnal}.

From our analysis into the different triads within our system, and their importance, we have identified elastic scattering and local interactions as the two dominant interactions once the system has sufficiently developed. In particular, our work is the first experimental evidence of elastic scattering as a significant type of triad interaction, corroborated by the extreme correlations of Fig. \ref{fig:bmd}. This is an important result given the complicated history of elastic scattering within the field. In the original work of McComas and Bretherton \cite{mccomas1977resonant}, elastic scattering was deemed responsible for attaining vertical symmetry in a GM spectrum, but was largely overlooked in favor of induced diffusion for the actual spectral power laws. More recent work investigating true power law solutions of the WKE have again found elastic scattering to be less significant than induced diffusion. Where our work assimilates best is with the results of Wu and Pan \cite{wu2023energy}, who found elastic scattering to be a significant nonlocal interaction. Further work on the importance of the individual nonlocal triads is required.

Our work also considered the contribution of local triads. Neglected in the McComas and Bretherton framework \cite{mccomas1977resonant}, the aforementioned studies of Lvov, Dematteis, and Pan \cite{dematteis2022origins,dematteis2023structure,pan2020numerical} have all drawn the conclusion that local interactions cannot be neglected in internal wave turbulence. Our results corroborate this idea. We found the vast majority of our interactions to be local. It is worth recalling that we arrived at this result under the same restrictive criteria as Wu and Pan \cite{wu2023energy}. Under more realistic definitions of scale separation, we would have an even greater contribution from local triads over nonlocal. Our work provides experimental validation to the analytical and numerical work done to promote local interactions. 

With this idea of triads in mind, it is worth hypothesizing a family tree of our waves. We determined that TRI is a significant interaction in our system during early times. Through instabilities, the number of distinct frequencies in the system slowly increases. At a certain point, TRI produces enough frequencies for other interactions to take further effect, and after that it appears that TRI is no longer significant. This may be in part due to a restriction on even lower frequencies. While our waves are well separated from the Coriolis frequency, it is possible that waves at lower frequencies cannot exist in our system. Considering Fig. \ref{fig:spatiotemporal}, by the time we get to the lowest frequencies we observe, the horizontal wavenumber is at the smallest value possible, which corresponds to a wavelength twice the length of the tank. If one of these waves were to try to engage in PSI/TRI and spawn daughter waves at even lower frequencies, they would have to be at a much larger $|m|$, since $|m/k|$ must increase but $|k|$ can decrease no further. The resulting waves could be at such small vertical length-scales that they are unstable to shear and immediately overturn. This is consistent with the idea of cutoff wavenumbers in the St. Laurent \cite{Laurent2002} and Kunze \cite{kunze2019unified} models, which for the ocean are around 10 m. Thus TRI can only go so far in our system, since at a certain point the waves it would try to create simply cannot survive in our domain. 

Once enough frequencies are born, local interactions and elastic scattering take over. Let's pause to speculate on how these triads affect the energy partitioning across scales. Wu and Pan \cite{wu2023energy} determined that while elastic scattering is very important for energy transfer across frequencies, it is relatively inefficient at moving energy across vertical wavenumbers. In contrast PSI and local interactions were found to do the heavy lifting in moving energy across vertical wavenumbers. By accommodating viscous effects in our classification, we identified a sequence of TRI as producing our gravest modes at $0.042\omega_0$ and $0.084\omega_0$. These waves were at particularly large values of $|m|$, which speaks to the cascade in vertical wavenumber brought about by TRI. However TRI is heavily discretized, and the few interactions we identified di not produce a sufficient distribution of $|m|$ to account for all of Fig. \ref{fig:spatiotemporal}(c). We suspect local triads help fill in the gaps in vertical wavenumbers, and lead to the development of a more continuous distribution of discrete vertical length scales. We identified more than 75\% of our triads as local, and given the range of vertical wavenumbers we expect these local interactions to be heavily involved in the vertical wavenumber cascade. 

As for elastic scattering, consider the frequencies $0.042\omega_0$ and $0.084\omega_0$ as the lower limit that $\omega$ can be. From that reasoning, any elastic scattering interaction involving these frequencies will necessarily involve an energy transfer toward higher frequencies, in increments of $0.042\omega_0$ and $0.084\omega_0$. What this means is that through elastic scattering we can fill up our ``continuum" of frequencies, by moving energy forward in tiny increments. So while elastic scattering may not do much to transfer energy in wavenumbers (as noted in \cite{wu2023energy}), we feel it is highly important in transferring energy amongst frequencies. This idea helps explain the conflicting ideas behind elastic scattering in the work of Dematteis \cite{dematteis2022origins, dematteis2023structure} and Wu and Pan \cite{wu2023energy}. The true power law solutions to the WKE are sans rotation. For a red spectrum one would expect elastic scattering to be most meaningful with the lowest frequency, which in this case would be $0$. The resulting interactions cannot move you forward in frequency, and without that, elastic scattering will have little effect. But if you do consider a nonzero lowest frequency, such as $f$, then elastic scattering by definition must move energy forward in increments of at least $f$. This was found by Wu and Pan \cite{wu2023energy}. In their study the Coriolis frequency was the lowest frequency possible, while in our work the ``lowest" wave frequencies were $0.042\omega_0$ and $0.084\omega_0$. Therefore, removing a lower limit on internal wave frequencies has the potential to undermine the importance of elastic scattering.

Additionally, we do not suspect our elastic scattering results to be biased based on preferential selection of what would otherwise be local interactions. Were we to tighten our definition of scale-separated interactions, the ease of finding elastic scattering in our system changes little. Since the primary culprits are two of the lowest frequencies in our tank, requiring a ratio $\omega_\mathrm{M}/\omega_\mathrm{L} \gg 2$ is still comfortably satisfied in Fig. \ref{fig:es}. Particularly for panel (b), we have over an order of magnitude separation between the lowest and middle frequency. As for the condition on the vertical wavenumbers, elastic scattering was first defined by McComas and Bretherton as having two high frequency wave at nearly equal and opposite vertical wavenumbers, and a low frequency wave at roughly double the vertical wavenumber \cite{mccomas1977resonant}. Thus expanding the regime of $|m_\mathrm{M}/m_\mathrm{H}|$ to go beyond roughly 0.5 would contradict the definition of elastic scattering itself. Therefore we feel justified in acknowledging its significance in our system. 

We should stress that our work, being limited to predominantly frequencies below $\omega_0$, cannot address questions on high frequency power laws, such as GM. By a similar vein, the lack of induced diffusion in our results should not question its role in oceanic transfer; the system matters considerably. To illustrate this point, in the numerical work of Pan \textit{et al.} \cite{pan2020numerical}, Figure 9 shows a bicoherence plot of their Hawaiin ridge simulations, which has a striking resemblance to our Fig. \ref{fig:bmd}. A low frequency wave seems to have an elevated correlation with roughly all other frequencies, much like in our system. But unlike in our results, Pan \textit{et al.} attribute these interactions to induced diffusion. This should not be seen as a contradiction between our work and theirs, but rather a reflection of the different domains and resulting waves. Under the greater freedom of $k$ and $m$ the ocean provides, and the greater separation between $\omega_0$ and $N$, induced diffusion is far more possible than in our confined tank. As mentioned earlier, in our setup elastic scattering and induced diffusion are somewhat mutually exclusive, given the limits on frequencies and wavenumbers. 

With that in mind, our research answers the following two questions, posed in the beginning of this work. First, we have demonstrated that wave turbulence can successfully move energy from tidal injection scales to dissipative scales, even absent a true forward cascade in frequency. This has been previously demonstrated numerically \cite{nikurashin2011mechanism, pan2020numerical}, but most experimental work on internal wave turbulence does not feature an internal tide forcing. We found a reduction in vertical scales by roughly a decade through wave-wave interactions. While we have not confirmed this, we suspect that the largest vertical wavenumbers seen represent a cutoff scale for waves in our system. Beyond this, the energy cascade is carried through some form of stratified and/or Kolmogorov turbulence. We will discuss the energy transfer beyond the wave regime in future work.  

Second, we have determined that within the realm of frequencies predominantly below the forcing, the fully developed system is composed predominantly of local triads and elastic scattering. Both of these results are important for the field, and add to the body of work highlighting the importance of local triads. We suspect that PSI/TRI is responsible for early energy transfer, and provides the necessary ingredients for other triads to take over. We reiterate though that experimental studies of WT are likely to fall into the viscous regime of TRI rather than true PSI. Further work unifying the concepts of TRI and nonlocal interactions is required to determine adequate distinctions under real-world conditions.  

From a global perspective, understanding the role internal waves play in mixing is an extremely important task to ensure accurate climate predictions. Currently climate models cannot resolve the internal wave driven dynamics, and rely on parameterizations \cite{mackinnon2013diapycnal, whalen2020internal}. But these parameterizations assume underlying physics, which must be justified. This issue is particularly relevant to the paradigm of research on climate change, since the effects of global warming have non-uniformly altered the abyssal stratification \cite{tan2023global}, and will continue to do so. The stratification sets the internal wave dynamics, and the wave dynamics in turn set the mixing \cite{kunze2019unified}. Thus if we want to understand how our oceans will behave years from now, we need to ascertain the underlying physics of the oceanic energy cascade.  

\section*{Acknowledgements}
This work is supported by NSF award OCE-2049213 MOMS: A Minimal Ocean Mixing System. Further support provided by the UNC Chapel Hill Fluids Lab. Special thanks to Jim Mahaney for aiding in the construction of the experiment, Bob Payne and Mathangi Mohanarajah for their aid in earlier experiments, and other members of the UNC Fluids Lab. Also special thanks to Yulin Pan for his consultation on some of the theoretical aspects of internal wave turbulence. 

\printbibliography

@article{wu2023energy,
	title        = {Energy cascade in the Garrett--Munk spectrum of internal gravity waves},
	author       = {Wu, Yue and Pan, Yulin},
	year         = 2023,
	journal      = {Journal of Fluid Mechanics},
	publisher    = {Cambridge University Press},
	volume       = 975,
	pages        = {A11}
}

@article{mccomas1977resonant,
	title        = {Resonant interaction of oceanic internal waves},
	author       = {McComas, C Henry and Bretherton, Francis P},
	year         = 1977,
	journal      = {Journal of Geophysical Research},
	publisher    = {Wiley Online Library},
	volume       = 82,
	number       = 9,
	pages        = {1397--1412}
}

@article{pan2020numerical,
	title        = {Numerical investigation of mechanisms underlying oceanic internal gravity wave power-law spectra},
	author       = {Pan, Yulin and Arbic, Brian K and Nelson, Arin D and Menemenlis, Dimitris and Peltier, WR and Xu, Wentao and Li, Ye},
	year         = 2020,
	journal      = {Journal of Physical Oceanography},
	volume       = 50,
	number       = 9,
	pages        = {2713--2733}
}

@article{dematteis2022origins,
	title        = {On the origins of the oceanic ultraviolet catastrophe},
	author       = {Dematteis, Giovanni and Polzin, Kurt and Lvov, Yuri V},
	year         = 2022,
	journal      = {Journal of Physical Oceanography},
	volume       = 52,
	number       = 4,
	pages        = {597--616}
}

@article{dematteis2023structure,
	title        = {The structure of energy fluxes in wave turbulence},
	author       = {Dematteis, Giovanni and Lvov, Yuri V},
	year         = 2023,
	journal      = {Journal of Fluid Mechanics},
	publisher    = {Cambridge University Press},
	volume       = 954,
	pages        = {A30}
}

@article{echeverri2009low,
	title        = {Low-mode internal tide generation by topography: an experimental and numerical investigation},
	author       = {Echeverri, Paula and Flynn, MR and Winters, Kraig B and Peacock, Thomas},
	year         = 2009,
	journal      = {Journal of Fluid Mechanics},
	publisher    = {Cambridge University Press},
	volume       = 636,
	pages        = {91--108}
}

@article{spiegel1960boussinesq,
	title        = {On the Boussinesq approximation for a compressible fluid.},
	author       = {Spiegel, Edward A and Veronis, G},
	year         = 1960,
	journal      = {Astrophysical Journal, vol. 131, p. 442},
	volume       = 131,
	pages        = 442
}

@article{aguilar2006laboratory,
	title        = {Laboratory generation of internal waves from sinusoidal topography},
	author       = {Aguilar, DA and Sutherland, BR and Muraki, DJ},
	year         = 2006,
	journal      = {Deep Sea Research Part II: Topical Studies in Oceanography},
	publisher    = {Elsevier},
	volume       = 53,
	number       = {1-2},
	pages        = {96--115}
}

@article{aguilar2006internal,
	title        = {Internal wave generation from rough topography},
	author       = {Aguilar, DA and Sutherland, BR},
	year         = 2006,
	journal      = {Physics of Fluids},
	publisher    = {AIP Publishing},
	volume       = 18,
	number       = 6
}

@article{lee2020evanescent,
	title        = {Evanescent to propagating internal waves in experiments, simulations, and linear theory},
	author       = {Lee, Allison and Hakes, Kyle and Liu, Yuxuan and Allshouse, Michael R and Crockett, Julie},
	year         = 2020,
	journal      = {Experiments in Fluids},
	publisher    = {Springer},
	volume       = 61,
	number       = 12,
	pages        = 252
}

@article{lee2019turning,
	title        = {Turning depths: Evanescent to propagating wave kinetic energy density},
	author       = {Lee, Allison and Crockett, Julie},
	year         = 2019,
	journal      = {Physical Review Fluids},
	publisher    = {APS},
	volume       = 4,
	number       = 3,
	pages        = {034803}
}

@article{rodda2022experimental,
	title        = {Experimental observations of internal wave turbulence transition in a stratified fluid},
	author       = {Rodda, Costanza and Savaro, Cl{\'e}ment and Davis, G{\'e}raldine and Reneuve, Jason and Augier, Pierre and Sommeria, Jo{\"e}l and Valran, Thomas and Viboud, Samuel and Mordant, Nicolas},
	year         = 2022,
	journal      = {Physical Review Fluids},
	publisher    = {APS},
	volume       = 7,
	number       = 9,
	pages        = {094802}
}

@article{davis2020succession,
	title        = {Succession of resonances to achieve internal wave turbulence},
	author       = {Davis, G{\'e}raldine and Jamin, Timoth{\'e}e and Deleuze, Julie and Joubaud, Sylvain and Dauxois, Thierry},
	year         = 2020,
	journal      = {Physical Review Letters},
	publisher    = {APS},
	volume       = 124,
	number       = 20,
	pages        = 204502
}

@article{lanchon2023internal,
	title        = {Internal wave turbulence in a stratified fluid with and without eigenmodes of the experimental domain},
	author       = {Lanchon, Nicolas and Mora, Daniel Odens and Monsalve, Eduardo and Cortet, Pierre-Philippe},
	year         = 2023,
	journal      = {Physical Review Fluids},
	publisher    = {APS},
	volume       = 8,
	number       = 5,
	pages        = {054802}
}

@article{van2020challenger,
	title        = {Challenger Deep internal wave turbulence events},
	author       = {van Haren, Hans},
	year         = 2020,
	journal      = {Deep Sea Research Part I: Oceanographic Research Papers},
	publisher    = {Elsevier},
	volume       = 165,
	pages        = 103400
}

@article{passaggia2020estimating,
	title        = {Estimating pressure and internal-wave flux from laboratory experiments in focusing internal waves},
	author       = {Passaggia, Pierre-Yves and Chalamalla, Vamsi K and Hurley, Matthew W and Scotti, Alberto and Santilli, Edward},
	year         = 2020,
	journal      = {Experiments in Fluids},
	publisher    = {Springer},
	volume       = 61,
	pages        = {1--29}
}

@article{camassa2018experimental,
	title        = {Experimental investigation of nonlinear internal waves in deep water with miscible fluids},
	author       = {Camassa, Roberto and Hurley, Matthew W and McLaughlin, Richard M and Passaggia, P-Y and Thomson, Colin FC},
	year         = 2018,
	journal      = {Journal of Ocean Engineering and Marine Energy},
	publisher    = {Springer},
	volume       = 4,
	pages        = {243--257}
}

@book{zakharov2012kolmogorov,
	title        = {Kolmogorov spectra of turbulence I: Wave turbulence},
	author       = {Zakharov, Vladimir E and L'vov, Victor S and Falkovich, Gregory},
	year         = 2012,
	publisher    = {Springer Science \& Business Media}
}

@article{lvov2001hamiltonian,
	title        = {Hamiltonian formalism and the Garrett-Munk spectrum of internal waves in the ocean},
	author       = {Lvov, Yuri V and Tabak, Esteban G},
	year         = 2001,
	journal      = {Physical review letters},
	publisher    = {APS},
	volume       = 87,
	number       = 16,
	pages        = 168501
}

@article{lvov2012resonant,
	title        = {Resonant and near-resonant internal wave interactions},
	author       = {Lvov, Yuri V and Polzin, Kurt L and Yokoyama, Naoto},
	year         = 2012,
	journal      = {Journal of Physical Oceanography},
	publisher    = {American Meteorological Society},
	volume       = 42,
	number       = 5,
	pages        = {669--691}
}

@article{hasselmann1962non,
	title        = {On the non-linear energy transfer in a gravity-wave spectrum Part 1. General theory},
	author       = {Hasselmann, Klaus},
	year         = 1962,
	journal      = {Journal of Fluid Mechanics},
	publisher    = {Cambridge University Press},
	volume       = 12,
	number       = 4,
	pages        = {481--500}
}

@book{nazarenko2011wave,
	title        = {Wave turbulence},
	author       = {Nazarenko, Sergey},
	year         = 2011,
	publisher    = {Springer},
	volume       = 825
}

@article{lvov2005scale,
	title        = {Scale Invariant Spectra of the Oceanic Internal Wave Field},
	author       = {Lvov, Yuri V and Polzin, Kurt L and Tabak, Esteban G},
	year         = 2005,
	journal      = {arXiv preprint math-ph/0505050},
	publisher    = {Citeseer}
}

@book{olbers1974energy,
	title        = {On the energy balance of small-scale internal waves in the deep-sea},
	author       = {Olbers, Dirk J{\"u}rgen},
	year         = 1974,
	publisher    = {GML Wittenborn},
	number       = 24
}

@article{olbers1976nonlinear,
	title        = {Nonlinear energy transfer and the energy balance of the internal wave field in the deep ocean},
	author       = {Olbers, Dirk J},
	year         = 1976,
	journal      = {Journal of Fluid mechanics},
	publisher    = {Cambridge University Press},
	volume       = 74,
	number       = 2,
	pages        = {375--399}
}

@article{caillol2000kinetic,
	title        = {Kinetic equations and stationary energy spectra of weakly nonlinear internal gravity waves},
	author       = {Caillol, Ph and Zeitlin, V},
	year         = 2000,
	journal      = {Dynamics of atmospheres and oceans},
	publisher    = {Elsevier},
	volume       = 32,
	number       = 2,
	pages        = {81--112}
}

@article{lvov2004hamiltonian,
	title        = {A Hamiltonian formulation for long internal waves},
	author       = {Lvov, Yuri and Tabak, Esteban G},
	year         = 2004,
	journal      = {Physica D: Nonlinear Phenomena},
	publisher    = {Elsevier},
	volume       = 195,
	number       = {1-2},
	pages        = {106--122}
}

@article{polzin2011toward,
	title        = {Toward regional characterizations of the oceanic internal wavefield},
	author       = {Polzin, Kurt L and Lvov, Yuri V},
	year         = 2011,
	journal      = {Reviews of geophysics},
	publisher    = {Wiley Online Library},
	volume       = 49,
	number       = 4
}

@article{wunsch2018100,
	title        = {100 years of the ocean general circulation},
	author       = {Wunsch, Carl and Ferrari, Raffaele},
	year         = 2018,
	journal      = {Meteorological Monographs},
	volume       = 59,
	pages        = {7--1}
}

@article{garrett1972space,
	title        = {Space-time scales of internal waves},
	author       = {Garrett, Christopher and Munk, Walter},
	year         = 1972,
	journal      = {Geophysical Fluid Dynamics},
	publisher    = {Taylor \& Francis},
	volume       = 3,
	number       = 3,
	pages        = {225--264}
}

@article{garrett1975space,
	title        = {Space-time scales of internal waves: A progress report},
	author       = {Garrett, Christopher and Munk, Walter},
	year         = 1975,
	journal      = {Journal of Geophysical Research},
	publisher    = {Wiley Online Library},
	volume       = 80,
	number       = 3,
	pages        = {291--297}
}

@mastersthesis{taebel2022investigation,
	title        = {Investigation of the Dual Cascade of Wave Turbulence in an Idealized Real-World System},
	author       = {Taebel, Zachary},
	year         = 2022,
	school       = {The University of North Carolina at Chapel Hill}
}

@book{cushman2011introduction,
	title        = {Introduction to geophysical fluid dynamics: physical and numerical aspects},
	author       = {Cushman-Roisin, Benoit and Beckers, Jean-Marie},
	year         = 2011,
	publisher    = {Academic press}
}

@article{allen1989statistical,
	title        = {A statistical mechanical explanation of the Garrett and Munk model of oceanic internal waves},
	author       = {Allen, KR and Joseph, RI},
	year         = 1989,
	journal      = {Johns Hopkins APL Technical Digest},
	publisher    = {JOHNS HOPKINS UNIV APPLIED PHYSICS LABORATORY ATTN: MANAGING EDITOR JOHN~\ldots{}},
	volume       = 10,
	number       = 4,
	pages        = {348--361}
}

@article{briscoe1975internal,
	title        = {Internal waves in the ocean},
	author       = {Briscoe, Melbourne G},
	year         = 1975,
	journal      = {Reviews of Geophysics},
	publisher    = {Wiley Online Library},
	volume       = 13,
	number       = 3,
	pages        = {591--598}
}

@article{alford2000observations,
	title        = {Observations of overturning in the thermocline: The context of ocean mixing},
	author       = {Alford, Matthew H and Pinkel, Robert},
	year         = 2000,
	journal      = {Journal of Physical Oceanography},
	publisher    = {American Meteorological Society},
	volume       = 30,
	number       = 5,
	pages        = {805--832}
}

@inproceedings{munk1966abyssal,
	title        = {Abyssal recipes},
	author       = {Munk, Walter H},
	year         = 1966,
	booktitle    = {Deep sea research and oceanographic abstracts},
	volume       = 13,
	number       = 4,
	pages        = {707--730},
	organization = {Elsevier}
}

@article{garrett2003internal,
	title        = {Internal tides and ocean mixing},
	author       = {Garrett, Chris},
	year         = 2003,
	journal      = {Science},
	publisher    = {American Association for the Advancement of Science},
	volume       = 301,
	number       = 5641,
	pages        = {1858--1859}
}

@article{wunsch2004vertical,
	title        = {Vertical mixing, energy, and the general circulation of the oceans},
	author       = {Wunsch, Carl and Ferrari, Raffaele},
	year         = 2004,
	journal      = {Annu. Rev. Fluid Mech.},
	publisher    = {Annual Reviews},
	volume       = 36,
	number       = 1,
	pages        = {281--314}
}

@article{garrett2007internal,
	title        = {Internal tide generation in the deep ocean},
	author       = {Garrett, Chris and Kunze, Eric},
	year         = 2007,
	journal      = {Annu. Rev. Fluid Mech.},
	publisher    = {Annual Reviews},
	volume       = 39,
	number       = 1,
	pages        = {57--87}
}

@article{egbert2000significant,
	title        = {Significant dissipation of tidal energy in the deep ocean inferred from satellite altimeter data},
	author       = {Egbert, Gary D and Ray, Richard D},
	year         = 2000,
	journal      = {Nature},
	publisher    = {Nature Publishing Group UK London},
	volume       = 405,
	number       = 6788,
	pages        = {775--778}
}

@article{whalen2020internal,
	title        = {Internal wave-driven mixing: Governing processes and consequences for climate},
	author       = {Whalen, Caitlin B and De Lavergne, Casimir and Naveira Garabato, Alberto C and Klymak, Jody M and MacKinnon, Jennifer A and Sheen, Katy L},
	year         = 2020,
	journal      = {Nature Reviews Earth \& Environment},
	publisher    = {Nature Publishing Group UK London},
	volume       = 1,
	number       = 11,
	pages        = {606--621}
}

@incollection{mackinnon2013diapycnal,
	title        = {Diapycnal mixing processes in the ocean interior},
	author       = {MacKinnon, Jennifer and St Laurent, Lou and Garabato, Alberto C Naveira},
	year         = 2013,
	booktitle    = {International Geophysics},
	publisher    = {Elsevier},
	volume       = 103,
	pages        = {159--183}
}

@article{Laurent2002,
	title        = {The role of internal tides in mixing the deep ocean},
	author       = {Laurent, L. S. and Garrett, C.},
	year         = 2002,
	journal      = {Journal of Physical Oceanography},
	volume       = 32,
	number       = 10,
	pages        = {2882--2899}
}

@article{StLaurent2002Tidal,
	title        = {Estimating tidally driven mixing in the deep ocean},
	author       = {St. Laurent, L. C. and Simmons, H. L. and Jayne, S. R.},
	year         = 2002,
	journal      = {Geophysical Research Letters},
	volume       = 29,
	number       = 23,
	pages        = {21--1}
}

@article{FoxKemper2019,
	title        = {Challenges and prospects in ocean circulation models},
	author       = {Fox-Kemper, B. and Adcroft, A. and B\"{o}ning, C. W. and Chassignet, E. P. and Curchitser, E. and Danabasoglu, G. and Eden, C. and England, M. H. and Gerdes, R. and Greatbatch, R. J. and Griffies, S. M.},
	year         = 2019,
	journal      = {Frontiers in Marine Science},
	volume       = 6,
	pages        = 65
}

@article{deLavergne2019,
	title        = {Toward global maps of internal tide energy sinks},
	author       = {de Lavergne, C. and Falahat, S. and Madec, G. and Roquet, F. and Nycander, J. and Vic, C.},
	year         = 2019,
	journal      = {Ocean Modelling},
	volume       = 137,
	pages        = {52--75}
}

@article{kunze2019unified,
	title        = {A unified model spectrum for anisotropic stratified and isotropic turbulence in the ocean and atmosphere},
	author       = {Kunze, Eric},
	year         = 2019,
	journal      = {Journal of Physical Oceanography},
	publisher    = {American Meteorological Society},
	volume       = 49,
	number       = 2,
	pages        = {385--407}
}

@article{staquet2002internal,
	title        = {Internal gravity waves: from instabilities to turbulence},
	author       = {Staquet, Chantal and Sommeria, Jo{\"e}l},
	year         = 2002,
	journal      = {Annual Review of Fluid Mechanics},
	publisher    = {Annual Reviews 4139 El Camino Way, PO Box 10139, Palo Alto, CA 94303-0139, USA},
	volume       = 34,
	number       = 1,
	pages        = {559--593}
}

@book{tanaka2022physics,
	title        = {Physics of Nonlinear Waves},
	author       = {Tanaka, Mitsuhiro},
	year         = 2022,
	publisher    = {Springer Nature}
}

@article{savaro2020generation,
	title        = {Generation of weakly nonlinear turbulence of internal gravity waves in the Coriolis facility},
	author       = {Savaro, Cl{\'e}ment and Campagne, Antoine and Linares, Miguel Calpe and Augier, Pierre and Sommeria, Jo{\"e}l and Valran, Thomas and Viboud, Samuel and Mordant, Nicolas},
	year         = 2020,
	journal      = {Physical Review Fluids},
	publisher    = {APS},
	volume       = 5,
	number       = 7,
	pages        = {073801}
}

@article{rodda2023internal,
	title        = {From internal waves to turbulence in a stably stratified fluid},
	author       = {Rodda, Costanza and Savaro, Cl{\'e}ment and Bouillaut, Vincent and Augier, Pierre and Sommeria, Jo{\"e}l and Valran, Thomas and Viboud, Samuel and Mordant, Nicolas},
	year         = 2023,
	journal      = {Physical Review Letters},
	publisher    = {APS},
	volume       = 131,
	number       = 26,
	pages        = 264101
}

@article{towne2018spectral,
	title        = {Spectral proper orthogonal decomposition and its relationship to dynamic mode decomposition and resolvent analysis},
	author       = {Towne, Aaron and Schmidt, Oliver T and Colonius, Tim},
	year         = 2018,
	journal      = {Journal of Fluid Mechanics},
	publisher    = {Cambridge University Press},
	volume       = 847,
	pages        = {821--867}
}

@article{mccomas1981time,
	title        = {Time scales of resonant interactions among oceanic internal waves},
	author       = {McComas, C Henry and M{\"u}ller, Peter},
	year         = 1981,
	journal      = {Journal of physical oceanography},
	volume       = 11,
	number       = 2,
	pages        = {139--147}
}

@article{xie2013observations,
	title        = {Observations of enhanced nonlinear instability in the surface reflection of internal tides},
	author       = {Xie, Xiaohui and Shang, Xiaodong and van Haren, Hans and Chen, Guiying},
	year         = 2013,
	journal      = {Geophysical Research Letters},
	publisher    = {Wiley Online Library},
	volume       = 40,
	number       = 8,
	pages        = {1580--1586}
}

@article{alford2008observations,
	title        = {Observations of parametric subharmonic instability of the diurnal internal tide in the South China Sea},
	author       = {Alford, MH},
	year         = 2008,
	journal      = {Geophysical Research Letters},
	publisher    = {Wiley Online Library},
	volume       = 35,
	number       = 15
}

@article{liu2020disintegration,
	title        = {Disintegration of the K 1 internal tide in the South China Sea due to parametric subharmonic instability},
	author       = {Liu, Kun and Zhao, Zhongxiang},
	year         = 2020,
	journal      = {Journal of Physical Oceanography},
	volume       = 50,
	number       = 12,
	pages        = {3605--3622}
}

@article{xie2011observations,
	title        = {Observations of parametric subharmonic instability-induced near-inertial waves equatorward of the critical diurnal latitude},
	author       = {Xie, Xiao-Hui and Shang, Xiao-Dong and van Haren, Hans and Chen, Gui-Ying and Zhang, Yuan-Zhi},
	year         = 2011,
	journal      = {Geophysical Research Letters},
	publisher    = {Wiley Online Library},
	volume       = 38,
	number       = 5
}

@article{nikurashin2011mechanism,
	title        = {A mechanism for local dissipation of internal tides generated at rough topography},
	author       = {Nikurashin, Maxim and Legg, Sonya},
	year         = 2011,
	journal      = {Journal of Physical Oceanography},
	publisher    = {American Meteorological Society},
	volume       = 41,
	number       = 2,
	pages        = {378--395}
}

@article{gayen2013degradation,
	title        = {Degradation of an internal wave beam by parametric subharmonic instability in an upper ocean pycnocline},
	author       = {Gayen, Bishakhdatta and Sarkar, S},
	year         = 2013,
	journal      = {Journal of Geophysical Research: Oceans},
	publisher    = {Wiley Online Library},
	volume       = 118,
	number       = 9,
	pages        = {4689--4698}
}

@article{wang2021resonant,
	title        = {On the resonant triad interaction over mid-ocean ridges},
	author       = {Wang, Shuya and Cao, Anzhou and Chen, Xu and Li, Qiang and Song, Jinbao},
	year         = 2021,
	journal      = {Ocean Modelling},
	publisher    = {Elsevier},
	volume       = 158,
	pages        = 101734
}

@article{richet2018internal,
	title        = {Internal tide dissipation at topography: triadic resonant instability equatorward and evanescent waves poleward of the critical latitude},
	author       = {Richet, O and Chomaz, J-M and Muller, C},
	year         = 2018,
	journal      = {Journal of Geophysical Research: Oceans},
	publisher    = {Wiley Online Library},
	volume       = 123,
	number       = 9,
	pages        = {6136--6155}
}

@article{mackinnon2005subtropical,
	title        = {Subtropical catastrophe: Significant loss of low-mode tidal energy at 28.9},
	author       = {MacKinnon, JA and Winters, KB},
	year         = 2005,
	journal      = {Geophysical Research Letters},
	publisher    = {Wiley Online Library},
	volume       = 32,
	number       = 15
}

@article{bourget2013experimental,
	title        = {Experimental study of parametric subharmonic instability for internal plane waves},
	author       = {Bourget, Baptiste and Dauxois, Thierry and Joubaud, Sylvain and Odier, Philippe},
	year         = 2013,
	journal      = {Journal of Fluid Mechanics},
	publisher    = {Cambridge University Press},
	volume       = 723,
	pages        = {1--20}
}

@article{bourget2014finite,
	title        = {Finite-size effects in parametric subharmonic{\'a}instability},
	author       = {Bourget, Baptiste and Scolan, H{\'U}l{\'R}ne and Dauxois, Thierry and Le Bars, Michael and Odier, Philippe and Joubaud, Sylvain},
	year         = 2014,
	journal      = {Journal of Fluid Mechanics},
	publisher    = {Cambridge University Press},
	volume       = 759,
	pages        = {739--750}
}

@article{joubaud2012experimental,
	title        = {Experimental parametric subharmonic instability in stratified fluids},
	author       = {Joubaud, Sylvain and Munroe, James and Odier, Philippe and Dauxois, Thierry},
	year         = 2012,
	journal      = {Physics of Fluids},
	publisher    = {AIP Publishing},
	volume       = 24,
	number       = 4
}

@article{clark2010generation,
	title        = {Generation, propagation, and breaking of an internal wave beam},
	author       = {Clark, Heather A and Sutherland, Bruce R},
	year         = 2010,
	journal      = {Physics of Fluids},
	publisher    = {AIP Publishing},
	volume       = 22,
	number       = 7
}

@article{sarkar2017topographic,
	title        = {From topographic internal gravity waves to turbulence},
	author       = {Sarkar, S and Scotti, A},
	year         = 2017,
	journal      = {Annual Review of Fluid Mechanics},
	publisher    = {Annual Reviews},
	volume       = 49,
	number       = 1,
	pages        = {195--220}
}

@article{dossmann2017mixing,
	title        = {Mixing and formation of layers by internal wave forcing},
	author       = {Dossmann, Yvan and Pollet, Florence and Odier, Philippe and Dauxois, Thierry},
	year         = 2017,
	journal      = {Journal of Geophysical Research: Oceans},
	publisher    = {Wiley Online Library},
	volume       = 122,
	number       = 12,
	pages        = {9906--9917}
}

@article{sutherland2013wave,
	title        = {The wave instability pathway to turbulence},
	author       = {Sutherland, Bruce R},
	year         = 2013,
	journal      = {Journal of Fluid Mechanics},
	publisher    = {Cambridge University Press},
	volume       = 724,
	pages        = {1--4}
}

@book{sutherland2010internal,
	title        = {Internal gravity waves},
	author       = {Sutherland, Bruce R},
	year         = 2010,
	publisher    = {Cambridge university press}
}

@article{Campagne2018,
	title        = {Impact of dissipation on the energy spectrum of experimental turbulence of gravity surface waves},
	author       = {Campagne, A. and Hassaini, R. and Redor, I. and Sommeria, J. and Valran, T. and Viboud, S. and Mordant, N.},
	year         = 2018,
	journal      = {Physical Review Fluids},
	volume       = 3,
	number       = 4,
	pages        = {044801}
}

@article{korobov2008interharmonics,
	title        = {Interharmonics in internal gravity waves generated by tide-topography interaction},
	author       = {Korobov, Alexander S and Lamb, Kevin G},
	year         = 2008,
	journal      = {Journal of Fluid Mechanics},
	publisher    = {Cambridge University Press},
	volume       = 611,
	pages        = {61--95}
}

@article{brouzet2016energy,
	title        = {Energy cascade in internal-wave attractors},
	author       = {Brouzet, Christophe and Ermanyuk, Evgeny V and Joubaud, Sylvain and Sibgatullin, Ilias and Dauxois, Thierry},
	year         = 2016,
	journal      = {Europhysics Letters},
	publisher    = {IOP Publishing},
	volume       = 113,
	number       = 4,
	pages        = 44001
}

@article{manders2004observations,
	title        = {Observations of internal tides in the Mozambique Channel},
	author       = {Manders, AMM and Maas, LRM and Gerkema, T},
	year         = 2004,
	journal      = {Journal of Geophysical Research: Oceans},
	publisher    = {Wiley Online Library},
	volume       = 109,
	number       = {C12}
}

@article{brouzet2016internal,
	title        = {Internal wave attractors examined using laboratory experiments and 3D numerical simulations},
	author       = {Brouzet, Christophe and Sibgatullin, IN and Scolan, Helene and Ermanyuk, EV and Dauxois, Thierry},
	year         = 2016,
	journal      = {Journal of Fluid Mechanics},
	publisher    = {Cambridge University Press},
	volume       = 793,
	pages        = {109--131}
}

@article{liao2012current,
	title        = {Current observations of internal tides and parametric subharmonic instability in Luzon Strait},
	author       = {Liao, Guanghong and Yuan, Yaochu and Yang, Chenghao and Chen, Hong and Wang, Huiqun and Huang, Weigen},
	year         = 2012,
	journal      = {Atmosphere-Ocean},
	publisher    = {Taylor \& Francis},
	volume       = 50,
	number       = {sup1},
	pages        = {59--76}
}

@article{chen2019can,
	title        = {Can tidal forcing alone generate a GM-like internal wave spectrum?},
	author       = {Chen, Zhiwu and Chen, Shaomin and Liu, Zhiyu and Xu, Jiexin and Xie, Jieshuo and He, Yinghui and Cai, Shuqun},
	year         = 2019,
	journal      = {Geophysical Research Letters},
	publisher    = {Wiley Online Library},
	volume       = 46,
	number       = 24,
	pages        = {14644--14652}
}

@article{tan2023global,
	title        = {On the global decrease in the deep and abyssal density stratification along the spreading pathways of Antarctic Bottom Water since the 1990s},
	author       = {Tan, Shuwen and Thurnherr, Andreas M},
	year         = 2023,
	journal      = {Geophysical Research Letters},
	publisher    = {Wiley Online Library},
	volume       = 50,
	number       = 11,
	pages        = {e2022GL102422}
}

@article{garcia2010robust,
	title        = {Robust smoothing of gridded data in one and higher dimensions with missing values},
	author       = {Garcia, Damien},
	year         = 2010,
	journal      = {Computational statistics \& data analysis},
	publisher    = {Elsevier},
	volume       = 54,
	number       = 4,
	pages        = {1167--1178}
}

@article{wang2012three,
	title        = {A three-dimensional gap filling method for large geophysical datasets: Application to global satellite soil moisture observations},
	author       = {Wang, Guojie and Garcia, Damien and Liu, Yi and De Jeu, Richard and Dolman, A Johannes},
	year         = 2012,
	journal      = {Environmental Modelling \& Software},
	publisher    = {Elsevier},
	volume       = 30,
	pages        = {139--142}
}

@article{oster1963density,
	title        = {Density Gradient Techniques.},
	author       = {Oster, Gerald and Yamamoto, Masahide},
	year         = 1963,
	journal      = {Chemical Reviews},
	publisher    = {ACS Publications},
	volume       = 63,
	number       = 3,
	pages        = {257--268}
}

\end{document}